\documentclass{aa}  
\usepackage{graphicx}
\usepackage{txfonts}
\usepackage{hyperref}

\usepackage[colorinlistoftodos]{todonotes} 

\usepackage{capt-of}
\usepackage{amsmath}
\usepackage{placeins}
\usepackage[utf8]{inputenc}
\usepackage{filecontents}
\usepackage{natbib}
\bibpunct{(}{)}{;}{a}{}{,} 
\usepackage{longtable} 
\usepackage{supertabular,booktabs} 

\begin{document} 

    \title{The tilt of the velocity ellipsoid in the Milky Way with \textit{Gaia} DR2}
    \titlerunning{The tilt of the velocity ellipsoid throughout the Milky Way}
    \author{Jorrit~H.~J.~Hagen\inst{\ref{inst1}}
    \and Amina~Helmi\inst{\ref{inst1}}
    \and P.~Tim~de~Zeeuw\inst{\ref{inst2},\ref{inst3}} 
    \and Lorenzo~Posti\inst{\ref{inst1},\ref{inst4}}
    } 
    \authorrunning{Hagen et al.}
    \institute{Kapteyn Astronomical Institute, University of Groningen,
              Landleven 12, 9747 AD Groningen, The Netherlands\\
              \email{hagen@astro.rug.nl}\label{inst1}
              \and Sterrewacht Leiden, Leiden University, Postbus 9513, 2300 RA Leiden, The Netherlands\label{inst2}
              \and Max Planck Institute for Extraterrestrial Physics, Giessenbachstrasse 1, 85748 Garching, Germany\label{inst3}
              \and Universit\'e de Strasbourg, CNRS UMR 7550, Observatoire astronomique de Strasbourg, 11 rue de l'Universit\'e, 67000 Strasbourg, France\label{inst4}
              }
    \date{Received 13 February 2019 / Accepted 07 August 2019}

 
    \abstract {The velocity distribution of stars is a sensitive probe
      of the gravitational potential of the Galaxy, and hence of its
      dark matter distribution. In particular, the shape of the dark
      halo (e.g. spherical, oblate, or prolate) determines velocity
      correlations, and different halo geometries are expected to
      result in measurable differences. Here we explore and interpret
      the correlations in the $(v_R, v_z)$-velocity distribution as a
      function of position in the Milky Way.  We selected a high-quality
      sample of stars from the \textit{Gaia} DR2 catalogue and
      characterised the orientation of the velocity distribution or
      tilt angle over a radial distance range of $[4-13]$~kpc and up
      to $3.5$~kpc away from the Galactic plane while taking into
      account the effects of the measurement errors. We find that the
      tilt angles change from spherical alignment in the inner Galaxy
      ($R\sim4$~kpc) towards more cylindrical alignments in the outer Galaxy ($R\sim11$~kpc) when using
      distances that take a global zero-point offset in
      the parallax of $-29\mu$as. However, if the amplitude of this
      offset is underestimated, then the inferred tilt angles in the
      outer Galaxy only appear shallower and are intrinsically more consistent with
      spherical alignment for an offset as large as $-54\mu$as.  We
      further find that the tilt angles do not seem to strongly vary
      with Galactic azimuth and that different stellar populations
      depict similar tilt angles. Therefore we introduce a simple
      analytic function that describes the trends found over the full
      radial range. Since the
      systematic parallax errors in {\it Gaia} DR2 depend on celestial
      position, magnitude, and colour in complex ways, it is not
      possible to fully correct for them. Therefore it will be
      particularly important for dynamical modelling of the Milky Way
      to thoroughly characterise the systematics in astrometry in
      future {\it Gaia} data releases.}


   \maketitle
%



\section{Introduction}
The second data release of the {\it Gaia} space mission
\citep{GaiaDR2_summary_Brownetal2018} contains more than $1.3$ billion
stars with measured proper motions and positions and a subset of over
$7$ million stars with full six-dimensional (6D) phase-space
information. The availability of the motions and positions of stars in
the Milky Way and its satellite galaxies has already led to new
insights about the Galaxy \citep[e.g.][]{Antojaetal2018, Belokurovetal2018, Helmietal2018,
Poggioetal2018_warp, Price-Whelan2018}, and many more discoveries
will likely follow before {\it Gaia}'s next data release.

Studies of the Galaxy provide insight about the formation and evolution
of galaxies in general, and hence about elements of the cosmological
paradigm. For example, detailed dynamical modelling of the Milky Way
and its satellites, and in particular their mass distribution,
provide critical constraints on the nature of dark matter
\citep[e.g.][]{Bonacaetal2018}. Mass models of the Galaxy, such as
those by \citet{McMillan2011}, \citet{Piffletal2014_constraininghalowithRAVE}, and \citet{McMillan2017}, have been developed to fit many different
observational constraints simultaneously, although this is very
challenging.  Therefore many works often focus on a specific aspect
such as the characterisation of the velocity
distribution across the Galaxy.

The in-plane velocity distribution $f(v_R,v_\phi)$ in the Solar
vicinity has long been known to be complex, and many moving groups are
known to exist \citep[e.g.][]{Proctor1869, Eggen1965, Dehnen1998,
  Antojaetal2008}. With \textit{Gaia} DR2 the level of detail visible
in the velocity distribution of stars has increased immensely
\citep[see e.g.][]{GaiaDR2_disckinematis_Katzetal2018,
  Antojaetal2018}, and a plethora of substructures have become
apparent. On the other hand, the 2D velocity distribution
describing the radial and vertical velocity components, $f(v_R,v_z)$, shows
significantly less substructure and the traditional velocity moments
can still describe the data well to first order.

Such velocity moments and thus the axial ratios of the velocity ellipsoid, however, depend on the stellar distribution function and are different for different populations of stars. In contrast, its orientation (or better known as alignment or tilt) is directly related to (the shape of) the underlying gravitational potential in which the stars move \citep[e.g.][]{vandeVenetal2003, Binney&Tremaine2008, Binney&McMillan2011, An&Evans2016} and is the focus of this paper.

Nearly spherically aligned velocity ellipsoids were found for the
halo \citep{Smithetal2009, Bondetal2010, Kingetal2015, Evansetal2016}
by mainly using data from the Sloan Digital Sky Survey
\citep{Yorketal2000}.  Similar findings were obtained by
\citet{Postietal2018} for dynamically selected nearby halo
stars. These authors obtained full 6D phase-space information by
combining radial velocity measurements from the RAdial Velocity
Experiment \citep[RAVE DR5,][]{Kunderetal2017} to the 5D subset of
the \textit{Gaia} DR1 catalogue \citep{GaiaDR1summary2016}. Most
recently, \citet{Weggetal2018_arxiv} used $15,651$ RR Lyrae halo stars
with accurate proper motions from \textit{Gaia} DR2 and also inferred
a nearly spherically aligned velocity ellipsoid over a large range of
distances between $1.5$~kpc and $20$~kpc from the Galactic centre.  When
fed into the Jeans equations, this result seems to imply a spherical
dark matter distribution.

Studies focusing on the orientation of the velocity ellipsoid in local
samples of the Milky Way disk have also been consistently reporting
(close to) spherical alignment. \citet{Siebertetal2008} have used RAVE
DR2 and found a tilt angle $\gamma$ equal to $7.3^\circ \pm 1.8^\circ$
for red clump stars at $R=R_\odot$ and $z=1$~kpc, where
$\gamma_\textrm{sph} = 7.1^\circ$ would be expected for spherical
alignment at this location. \citet{Casettietal2011} found
$8.6^\circ \pm 1.8^\circ$ for a sample of stars with heights between
$0.7$~kpc and $2.0$~kpc and representative of the metal-rich thick disk,
which can be compared to $\gamma_\textrm{sph} = 8.0^\circ$ given the mean location of the
sample.  Subsequently, \citet{Smithetal2012} reinforced these findings
using data from the Sloan Digital
Sky Survey DR7 \citep[SDSS,][]{Abazajian2009}. 
\citet{Binneyetal2014} using RAVE data, and \citet{Budenbenderetal2015}, using Sloan
Extension for Galactic Understanding and Exploration
\citep[SEGUE,][]{Yannyetal2009}, characterised the tilt angle around
the Galactic radius of the Sun up to $z\sim2.0$~kpc by
$\gamma(z) \approx a_0 \arctan(z/R_\odot)$. They found $a_0\sim 0.8$
and $a_0=0.9 \pm 0.04$ respectively, values close to, but
significantly different from, spherical alignment for which $a_0$ = 1.0.
Recently, \citet{Mackerethetal2019_arxiv} have analysed the kinematics
of mono-age, mono-[Fe/H] populations for both low and high
[$\alpha$/Fe] samples.  They have cross matched the Apache Point
Observatory Galactic Evolution Experiment \citep[APOGEE
DR14,][]{Majewskietal2017} with Gaia DR2 to obtain a sample of
$65,719$ red giant stars located between $4$~kpc and $13$~kpc in
Galactic radius and up to $2$~kpc from the Galactic plane.
\citet{Mackerethetal2019_arxiv} report that the tilt angles found are
consistent with spherical alignment for all populations, although they
note that the uncertainties are very large.

In this work we characterise the orientation of the velocity ellipsoid
over a larger section of the Milky Way by using a dataset of
more than 5 million stars from \textit{Gaia} DR2.  The paper is
organised as follows.  In Sect. \ref{sec:data} the dataset is
introduced as well as the selection criteria applied. In
Sect. \ref{sec:methods} we characterise the velocity distribution and
the measurement errors. 
The results are presented in Sect. \ref{sec:results}. In that section
we also explore differences with azimuth, investigate trends with
stellar populations, and put forward a fit that reproduces the
variation of the tilt angle with position in the Galaxy. In
Sect. \ref{sec:discussion} we explore the effect of systematic errors
on our measurements and show that the systematic parallax errors present 
in \textit{Gaia} DR2 have a significant impact on the tilt angles found.
In that Section we therefore also discuss our findings in the context of
Galactic models. We
summarise in Sect. \ref{sec:conclusions}.




\section{Data}
\label{sec:data}

We used the subset of {\it Gaia} DR2 with full 6D information
\citep{GaiaDR2_disckinematis_Katzetal2018}. We use the Bayesian distance 
estimates $\hat{d}$ provided by \citet{McMillan2018_simpleGaiaDR2distances_RVSsubset} who 
uses the \textit{Gaia} DR2 parallaxes $\varpi$ and $G_{\textrm{RVS}}$ magnitudes as input. \citet{McMillan2018_simpleGaiaDR2distances_RVSsubset} takes into account
\textit{Gaia} DR2's overall parallax offset of $-29 \mu$as with a RMS error of $43 \mu$as \citep{GaiaDR2_astrometricsolution_Lindegren2018}. 

To construct a high-quality sample we select
stars with at most 20\% relative distance
errors, that is $\hat{d}/\hat{\epsilon}(\hat{d})>5$, and $\hat{d}<5$~kpc.
The sample contains $5,796,226$ stars. Stars with $\hat{d}<1$~kpc, typically have
distances better than 5\% (median 2.8\%) and for stars at $4<\hat{d}<5$~kpc
the relative distance errors are in between 12\% and 20\% (median
17.1\%).

\begin{figure}[t] \centering 
\includegraphics[width=0.5\textwidth]{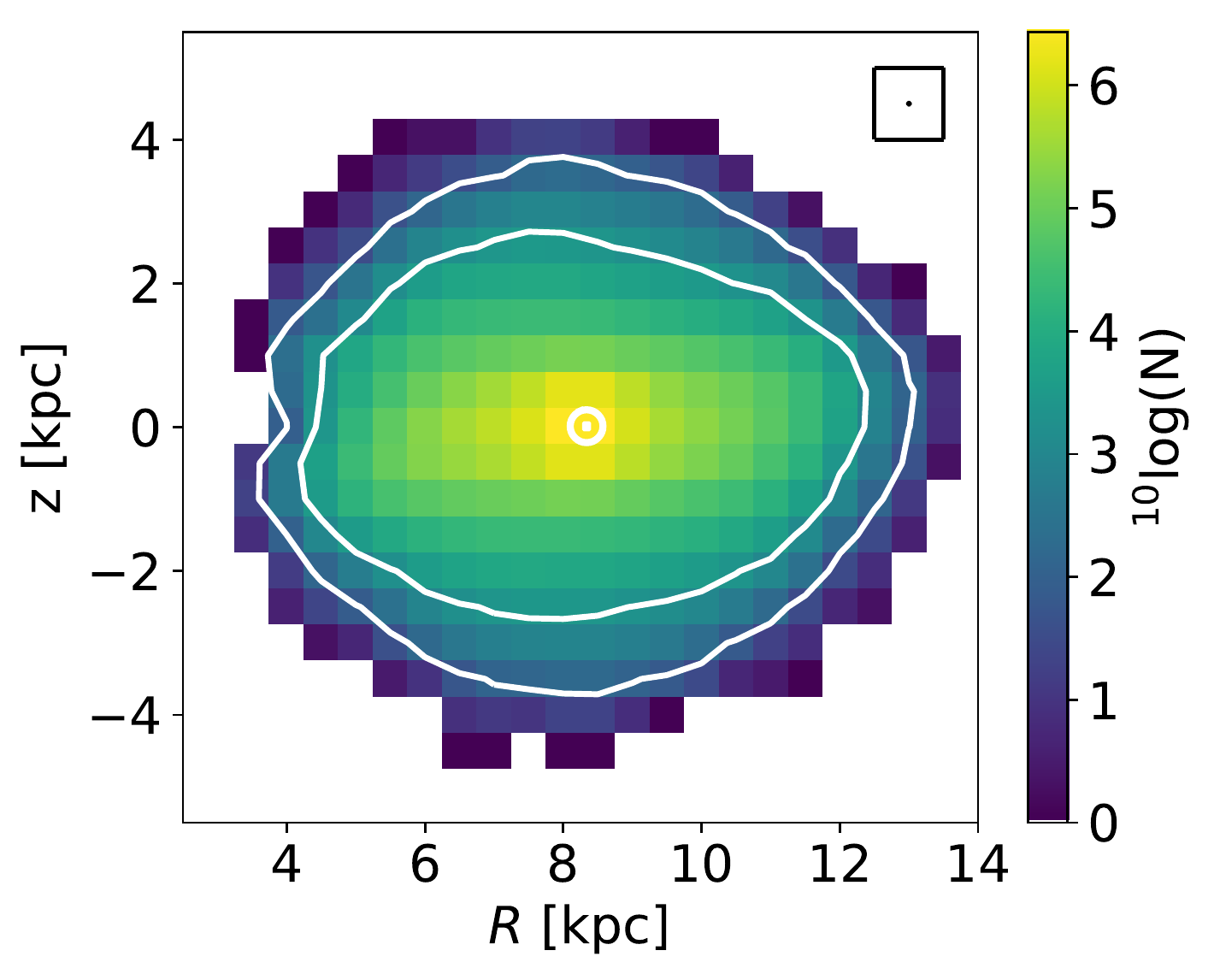} 
\caption{Star counts from our high-quality \textit{Gaia} DR2 6D sample in bins of width $1.0$~kpc in $R$ and $z$, as indicated by the box in the upper right corner. The central coordinates of the bins are separated by $0.5$~kpc in $R$ and $z$, thus the bins are not fully independent. The white contours indicate the location of bins with $2,000$ (inner contour) or $100$ (outer contour) stars. The position of the Sun is indicated by the white symbol. 
Only stars with $\hat{d}/\hat{\epsilon}(\hat{d})>5$ are considered in our sample.} 
\label{fig:Nstars_2d} 
\end{figure}

In Fig. \ref{fig:Nstars_2d} we show the extent of our sample in a number
density map. To compute the Galactocentric cylindrical coordinates
$(R,z,\phi)$, we assume\footnote{Use of the value of $R_\odot = 8178\pm13_{stat.}\pm22_{sys.}$ pc, as determined by \citet{Gravityetal2019_Rsun}, does not affect the main conclusions of this paper.} $R_\odot = 8.3$~\text{kpc}
\citep{Schonrich2012} and $z_\odot = 0.014$~\text{kpc}
\citep[][and $\phi_\odot = 180^\circ$]{Binneyetal1997}  for the position
of the Sun. Because of the imposed maximum distances to the stars, the
sample extends from $R\sim4$~kpc up to $R\sim13$~kpc and reaches up to
$z=\pm4$~kpc. The white contours in Fig.\ref{fig:Nstars_2d} indicate
the location of bins containing $2,000$ and $100$ stars
respectively. This shows that Galactic heights up to $\sim 3.5$~kpc are
still covered with a statistically significant number of stars.

We derive the velocities of the stars in our sample in a
Galactocentric cylindrical coordinate system ($v_R$, $v_z$, $v_\phi$).
For the motion of the Local Standard of Rest (LSR), that is the velocity of a circular orbit at $R=R_\odot$, 
we assume $v_\mathrm{c}(R_\odot) = 240$~\text{km/s}
\citep[][]{Piffletal2014_constraininghalowithRAVE, Reidetal2014}.
The peculiar motion of the Sun with respect to the LSR is taken to be $(U,V,W)_\odot = (11.1, 12.24, 7.25)$~km/s \citep{Schonrichetal2010}, where $U$ denotes
motion radially inwards and $V$ in the direction of Galactic rotation
(both in the Galactic plane), and $W$ perpendicular to the Galactic
plane and in the direction of the Galactic north pole. 
We propagate the errors and correlations in the observables to determine
the errors on the velocities (and their correlations). Here we assume that the Bayesian distances
are not correlated with the remaining astrometric parameters.
The velocity errors for the stars in our sample at $\hat{d}<1$~kpc are typically smaller than $2$~km/s with a median value of $\sim 1$~km/s for the $v_R$-, $v_z$-, and $v_\phi$-components. At $4<\hat{d}<5$~kpc the median errors are in the range from $\sim3$~km/s to $\sim8$~km/s and generally smaller than $15$~km/s.

The characterisation of the kinematics, in terms of the mean motions
and velocity dispersions, of a large part of the Milky Way disk have
been presented in \citet{GaiaDR2_disckinematis_Katzetal2018} using the
6D dataset from \textit{Gaia} DR2. This characterisation has put on
firm ground the evidence of the presence of streaming motions in all
velocity components \citep{Siebertetal2008, Williamsetal2013,
  Tianetal2017, Carrilloetal2018} and revealed a large amount of
substructure in the velocity distributions. In this paper we proceed to focus on the correlation between the radial and vertical
velocity components across a large fraction of the Milky Way galaxy.




\section{Methods}
\label{sec:methods}

The 3D velocity distribution of stars $f(v_\phi,v_R, v_z)$ at a given
point in the Galaxy may be characterised by its various moments. As
described in the Introduction, the tilt of the velocity ellipsoid
refers to the orientation of the 2D velocity distribution
$f(v_R,v_z)$, which would be obtained by integrating over $v_\phi$.
As shown in \citet{Smithetal2009} and \citet{Budenbenderetal2015},
this is equivalent to taking the moments of the 3D velocity
distribution and neglecting the cross terms with $v_\phi$. These cross-terms
are interesting in their own right, as they reveal also other physical
mechanisms at work, such as for example the presence of substructures
associated to resonances induced by the rotating Galactic bar
\citep{Dehnen1998}, but are not the focus of this work.



\subsection{The tilt angle: the orientation of the velocity ellipse}
\label{subsec:tilt}

In the Galactocentric cylindrical coordinate system we define the tilt angle $\gamma$, following for instance \citet{Smithetal2009}, as:
\begin{equation}
\label{eq:tiltangle}
    \tan(2 \gamma) = \frac{2 \mathrm{cov}(v_R, v_z)}{\mathrm{var}(v_R) - \mathrm{var}(v_z)} \, ,
\end{equation}
which therefore takes values from $-45$ degrees to $+45$ degrees, and is
measured counterclockwise (i.e. from the $v_R$-axis towards the positive
$v_z$-axis). For exact cylindrical alignment $\gamma_{\textrm{cyl}}=0^\circ$ 
and the major and minor axis align
with the Galactocentric cylindrical coordinates.

It is also possible to define a tilt angle $\alpha$ with respect to the spherical coordinate system $(r,\theta,\phi)$, where $\tan(\theta) \equiv R/z$, that is:
\begin{equation}
\label{eq:tiltangle_sph}
    \tan(2 \alpha) \equiv \frac{2 \mathrm{cov}(v_r, v_\theta)}{\mathrm{var}(v_r) - \mathrm{var}(v_\theta)} \, .
\end{equation} 
The tilt angle $\alpha$ thus measures directly the deviation from
spherical alignment, which corresponds to $\alpha = 0^\circ$. In such a case one of the
principal axes of the ellipse points to the Galactic centre.
The relation
between $\alpha$ and $\gamma$ at every $(R,z)$ is
\begin{equation}
    \tan(2\gamma) = - \tan(2\theta + 2\alpha).
\label{eq:tilt-relations}
\end{equation}

From now on, we always refer to the tilt angle $\gamma$, thus as
defined in the cylindrical coordinate system, unless stated otherwise.
To explore the spatial variation of the tilt angle we measure the
intrinsic moments of Eq.~\ref{eq:tiltangle} after projecting all stars
onto the $(R,z)$-plane, thus ignoring in the first stage the Galactic
azimuthal angle of the stars (although this is considered in 
Sect. \ref{subsec:releasingaxisymmetry}). We bin the meridional plane
as in Fig. \ref{fig:Nstars_2d} and always require at least $100$ stars
per bin.



\subsection{Accounting for measurement errors}
\label{subsec:errordeconvolution}

Measurement errors affect the observed velocity moments and can therefore have a significant effect on the inferred tilt angles \citep{Siebertetal2008}.
To establish their effect we here explore two `methods' to
account for the errors and for recovering the (intrinsic) velocity moments.

Method 1. We assume that the stars in a given spatial bin have similar
measurement errors. This assumption is reasonable because the
measurement errors in a particular bin are usually much smaller than
the intrinsic velocity dispersion. If the measurement errors were
exactly the same for all stars in a bin, the intrinsic velocity
covariance matrix can be recovered by subtracting the error covariance
matrix from the observed velocity covariance matrix. This follows from
the fact that convolving a Gaussian distribution with Gaussian
distributed measurement errors again results in a Gaussian with
covariance matrix
$\vec{\Sigma_{\mathrm{obs}}} = \vec{\Sigma_{\mathrm{intr}}} +
\vec{\Sigma_{\mathrm{error}}}$, where $\vec{\Sigma_{\mathrm{obs}}}$
and $\vec{\Sigma_{\mathrm{intr}}}$ are the observed and intrinsic
covariance matrix of the velocity distribution respectively. In our
approximation
$\vec{\Sigma_{\mathrm{error}}} \approx \mathrm{median}\left(
  \vec{\Sigma_{\mathrm{error,i}}}\right)$ for
\begin{equation}
\vec{\Sigma_{\mathrm{error,i}}} =
\begin{bmatrix}
     \mathrm{var}(v_{R,i})              & \mathrm{cov}(v_{R,i}, v_{z,i}) \\
     \mathrm{cov}(v_{R,i}, v_{z,i})      & \mathrm{var}(v_{z,i})
\end{bmatrix} \, ,
\end{equation}
in which the diagonal terms denote the variance error of the corresponding velocity component of star $i$. Similarly $\mathrm{cov}(v_{R,i}, v_{z,i})$ denotes the error covariance for the $(v_{R}, v_{z})$ measurements of star $i$. For the required typical errors we take the relevant median errors of the stars in the bin. The recovered intrinsic velocity moments are then used to characterise the velocity distribution. The errors on these moments are analytically estimated and then propagated into uncertainties on the recovered tilt angles. More details can be found in Appendix \ref{App:Appendix_errorsofmoments}.

Method 2. We perform Markov Chain Monte Carlo (MCMC) modelling
\citep{Foreman-MacKeyetal2013} for bins with a smaller number of stars
(with $100<N<2,000$). This aims to solve for the intrinsic velocity
dispersions $\sigma(v_R)_\textrm{intr}$ and
$\sigma(v_z)_\textrm{intr}$, the mean velocities $\langle v_R \rangle$
and $\langle v_z \rangle$, and the covariance term
$\textrm{cov}(v_R, v_z)_\textrm{intr}$ in each bin. This is done by
maximizing the bivariate Gaussian likelihood function
$L=\prod^{N}_{i=1} L_i$, where
\begin{equation}
\begin{split}
L_i &= L_i[\langle v_R \rangle, \sigma(v_R)_\textrm{intr}, \langle v_z \rangle, \sigma(v_z)_\textrm{intr}, \textrm{cov}(v_R, v_z)_\textrm{intr}] \\
& = \frac{1}{\sqrt{\textrm{det}(2 \pi \vec{\Sigma_i})}} \exp \left[
-\frac{1}{2} (\vec{x_i} -\vec{\mu})^\intercal \vec{\Sigma_i}^{-1} (\vec{x_i}-\vec{\mu})
\right] \, ,
\end{split}    
\label{eq:likelihood}
\end{equation}
in which $\vec{x_i} = [v_{R,i}, v_{z,i}]$, $\vec{\mu} = [\langle v_R \rangle, \langle v_z \rangle]$ and $\vec{\Sigma_i} = \vec{\Sigma_\mathrm{intr}} + \vec{\Sigma_{\mathrm{error},i}}$. Whereas in Method 1 $\vec{\Sigma_{\mathrm{error},i}}$ was assumed to be the same for each star, we here use $\vec{\Sigma_{\mathrm{error},i}}$ for each star separately. We add priors to the model that only allow for positive velocity dispersions in $v_R$ and $v_z$ and that restrict the correlation coefficient between $v_R$ and $v_z$ always to be within [$-1$,$1$]. For a given bin, the samples drawn by the MCMC run translate into a distribution of tilt angles. We take the median as the best estimate of the tilt angle. For its error we take half the difference between the tilt angles corresponding to the $16^\mathrm{th}$ and $84^\mathrm{th}$ percentile. \\

In general we find that the effect of the measurement errors on the
recovered moments is small. Moreover, for most bins the velocity
measurement errors are sufficiently similar and small that we may use
the computationally much faster Method 1 instead of the
MCMC-based deconvolution. We have also compared the results to the
case in which we simply compute the variances of the observed stellar
velocities in the bins of interest, and take these at face value,
meaning that we do not take into account the measurement errors. The results
are again rather similar, see for example, Fig.~\ref{fig:exampleerror}
of Appendix~\ref{App:Appendix_errorsofmoments} which shows the
distributions of the measurement errors for the bin located at
$R=11.5$~kpc and $z=1.5$~kpc. In what follows, we use the
results from Method 1 unless stated otherwise.




\section{Results}
\label{sec:results}

\begin{figure*}[t!]
  \centering
  \includegraphics[width=0.7\textwidth]{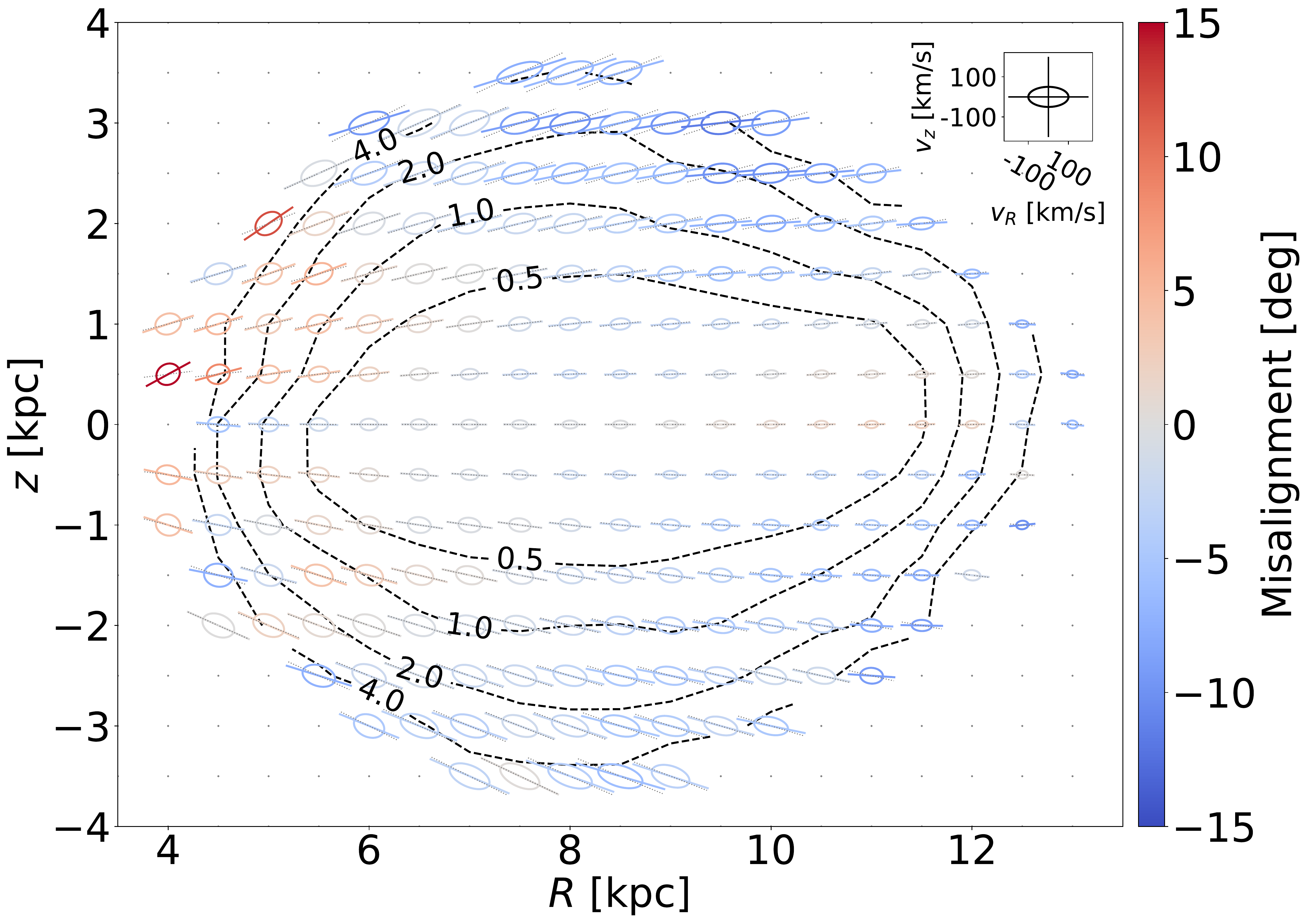}%
  \caption{Velocity ellipses in the meridional plane. The ellipses are colour-coded by their misalignment with respect to spherical alignment. The orientation that corresponds to spherical alignment is indicated by the dotted grey line through each ellipse. The inset in the top right of the figure shows the velocity ellipse for a non-tilted distribution with dispersions $\sigma(v_R)=100$~km/s and $\sigma(v_z)=50$~km/s (see Sect. \ref{sec:results} for more information). The contours show the (relatively small) formal statistical errors on the recovered tilt angles and are drawn for error levels of [$0.5,1.0,2.0,4.0$] degrees. 
  See Sect. \ref{sec:discussion} for a discussion on the effect of systematic errors.} 
  \label{fig:misalignment} 
\end{figure*}%

We present our measurement of the tilt angles by showing velocity
ellipses in the meridional plane.  At each position $(R,z)$, we define
a set of axes with $v_R$ into the $R$-direction and $v_z$ in the
$z$-direction. The centre of each velocity ellipse is always placed at
its position $(R,z)$. The size of the major and minor axis of each
ellipse scale with the intrinsic velocity dispersions along these
directions. The $R$- and $z$-axis are both scaled by the same constant
$c_x$. Similarly, all $v_R$- and $v_z$-axes are scaled by a constant
$c_v$, thus both sets of axes have an aspect ratio of $1$. As a
consequence, the velocity ellipses drawn will actually point to the
Galactic centre when there is spherical alignment. As a reference, the
inset in the figures shows a velocity distribution aligned in 
cylindrical coordinates and with $\sigma(v_R)=100$~km/s and
$\sigma(v_z)=50$~km/s (unless stated otherwise).


\subsection{Tilt angles projected onto the $(R,z)$-plane}
\label{subsec:axisymmetry}

Fig. \ref{fig:misalignment} shows the velocity ellipses colour-coded
by their angular misalignment with respect to spherical alignment. For
$z\geq0$~kpc we define this misalignment as
$\gamma - \gamma_{\textrm{sph}}$, whereas for $z<0$~kpc the
misalignment is $\gamma_{\textrm{sph}} - \gamma$. Steeper tilt angles
result in positive misalignment (from light to dark red), shallower tilt
angles in negative misalignment (from light to dark blue). Ellipses that 
are consistent with spherical alignment are greyish. At the midplane it is
however not possible to distinguish between spherical and cylindrical alignment, since 
both $\gamma_{\textrm{sph}} = \gamma_{\textrm{cyl}}=0^\circ$ at $z=0$~kpc, thus here consistency with spherical alignment also implies consistency with cylindrical alignment. Only away from the midplane it is possible to differentiate between these types of alignment.

We further add contours of constant formal statistical error values on the recovered tilt angles in Fig.~\ref{fig:misalignment}. We have drawn contours for errors reaching $0.5$, $1.0$, $2.0$, and $4.0$ degrees. These contours show the great quality of our dataset over the distance range explored.

From this figure it is evident that there are
just a few bins that have tilt angles much steeper than spherical
alignment (i.e. there are just two dark red ellipses). These are however located in the 
inner regions of the Galaxy and at those positions where the error on the tilt angle is also large.

In general, however, the following trend is apparent: for Galactocentric spherical radius $r \sim 4$~kpc, the orientations of the velocity ellipses seem to be slightly steeper than spherical alignment. For $r \sim 7$~kpc they seem fully consistent with spherical alignment. For larger radii, that is $R>8$~kpc and $|z| \gtrsim 1$~kpc, the ellipses have a negative misalignment, meaning that the orientations of the ellipses become shallower compared to prediction for spherical alignment. Here the orientation thus changes into the direction of cylindrical alignment and is no longer consistent with spherical alignment. 

\begin{figure*}[t]
  \centering
   \includegraphics[width=1.0\textwidth]{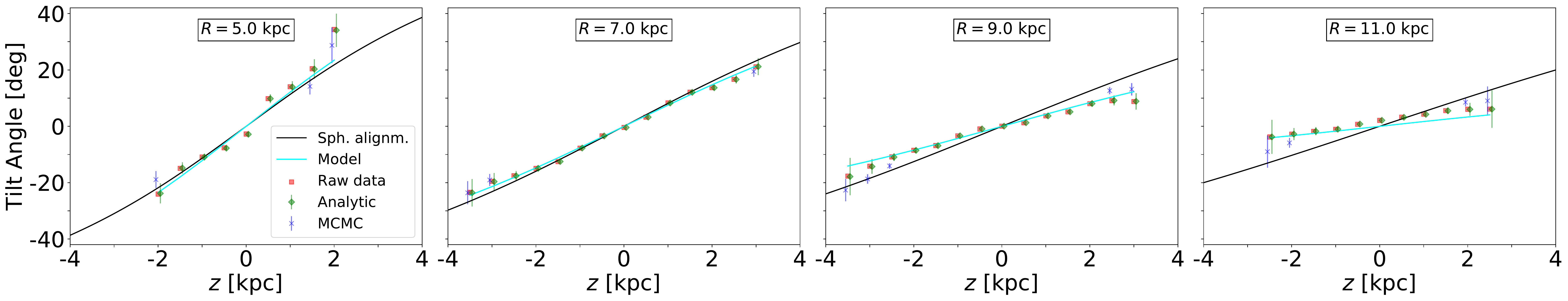}
  \caption{Tilt angles as a function of Galactic height for different
    positions across the Galaxy. We show the trends with $z$ for
    $R=[5,7,9,11]$~kpc. The red squares, green diamonds, and blue crosses 
    are based on the methods described in
    Sect. \ref{subsec:errordeconvolution} (see text). The solid black line shows the trend that would correspond to spherical alignment. The
    tilt angle is changing from spherical alignment in the
    inner Galaxy ($R \sim 5$~kpc) towards shallower tilt angles 
    at $R \sim 11$~kpc. 
    The cyan line shows the analytic description of the data as proposed in Sect. \ref{subsec:quantifying}. 
    }
  \label{fig:exampleradii} 
\end{figure*}%

To be able to assess whether the tilt angles found are more consistent with spherical or cylindrical alignment we show them with error bars in Fig. \ref{fig:exampleradii} as a function of height for four Galactic radii, namely $R=[5,7,9,11]$~kpc. The red squares (without error bars; labelled `Raw data') follow from
computing the moments directly from the data, and the green diamonds (`Analytic') and blue crosses (`MCMC') are derived using Method 1 and Method 2 respectively, thus accounting for the
measurement errors (see Sect.~\ref{subsec:errordeconvolution}). They give consistent results given the error bars, although the MCMC-method seems to result in slightly steeper tilt angles.

The black curve in Fig. \ref{fig:exampleradii} shows the expectation
in the case of spherical alignment. At $R=5$~kpc
(left panel) the recovered tilt angles are in agreement with
spherical alignment for the heights explored.  At $R=7$~kpc (left
centre panel) the data is consistent with spherical alignment up
to $|z|\sim2$~kpc. For larger heights the tilt angles are only mildly shallower.
For $R=9$~kpc and $R=11$~kpc, however, the tilt angles are becoming increasingly shallower with respect to spherical alignment. In fact, for $R=12$~kpc (see Fig. \ref{fig:misalignment}) the orientation of the ellipses become more consistent with cylindrical alignment for the heights probed.

\begin{figure}
  \centering
  \includegraphics[width=0.5\textwidth]{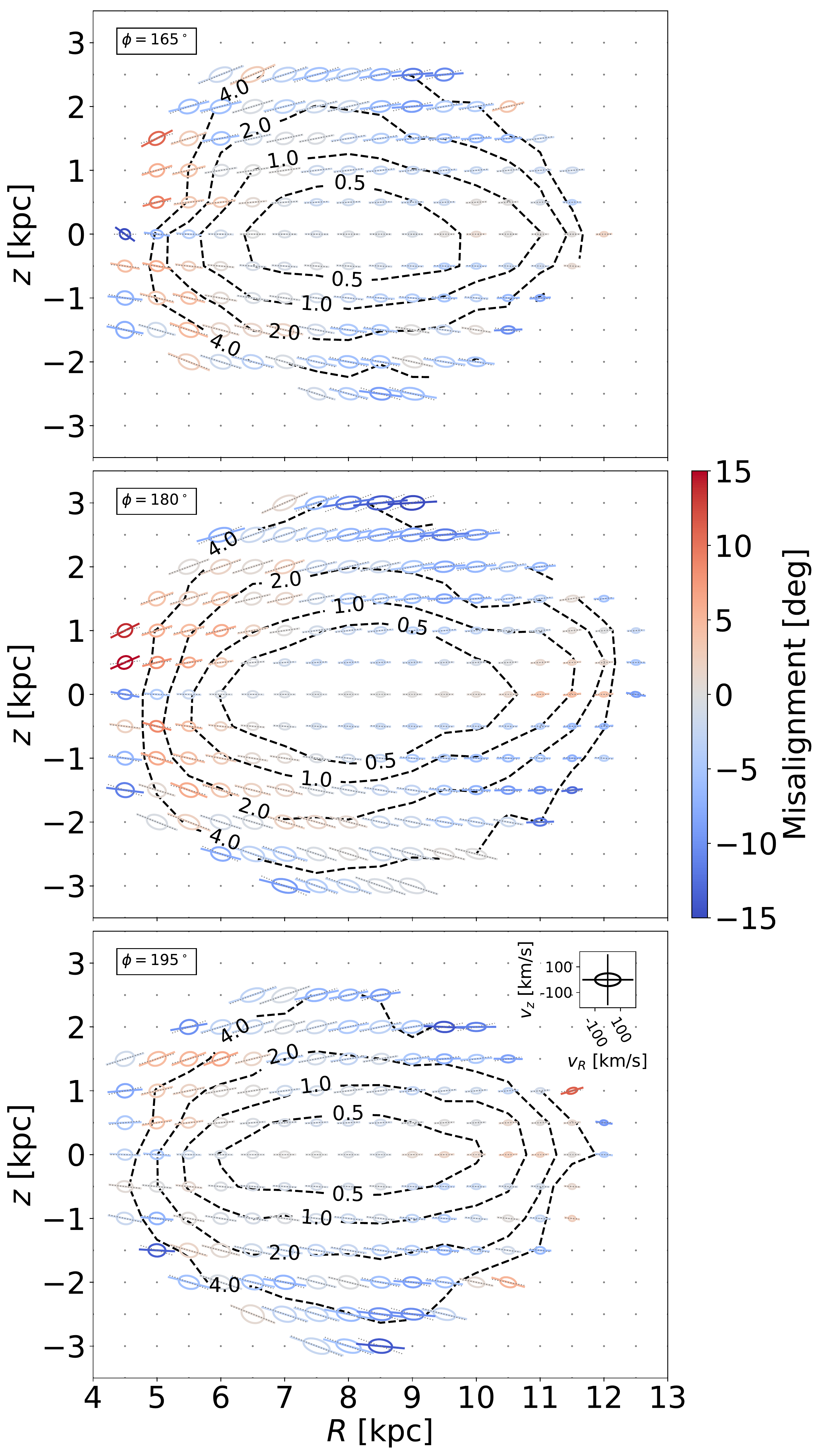}
  \caption{Velocity ellipses in the meridional plane, now for different positions in azimuth ($\phi = [165^{\circ}, 180^{\circ},195^{\circ}]$ from top to bottom, respectively). The spatial bins are cubes in $(x,y,z)$, of $1$~kpc on a side. The colours of the ellipses represent the misalignment with respect to spherical alignment (as in Fig. \ref{fig:misalignment}).} 
  \label{fig:tilt_difazimuths} 
\end{figure}%


\subsection{Tilt angles for different azimuthal angles}
\label{subsec:releasingaxisymmetry}

\begin{figure*}[t]
  \centering
  \includegraphics[width=\textwidth]{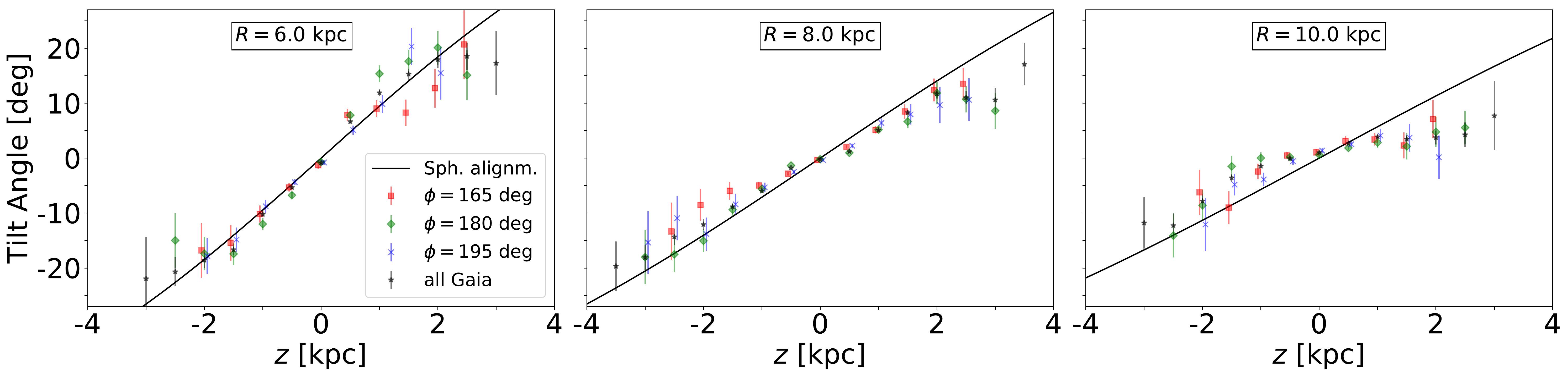}
  \caption{Tilt angles as a function of Galactic height for different radial and azimuthal positions across the Galaxy. The red squares, green diamonds, and blue crosses show the measurements for $\phi=[165^\circ,180^\circ,195^\circ]$, respectively. The black starred symbols show the measurements irrespective of azimuth (as in Sect. \ref{subsec:axisymmetry}). Given the error bars, there are only small differences in the tilt angles for the different azimuths explored. The solid black line denotes the trend expected for spherical alignment.} 
  \label{fig:exampleradii_difphi} 
\end{figure*}%

Since the Galaxy is not axisymmetric we now investigate whether the
tilt angles vary with azimuth by taking into account the 3D location
of the individual stars in our dataset. We bin the data into Cartesian
bins $(x,y,z)$ whose volume is fixed to 1$\times$1$\times$1~kpc$^3$,
which implies that the different azimuthal cones we explore 
contain independent data for $R>4$~kpc. These cones are centred on
three different angles $\phi = [165^{\circ},180^{\circ}, 195^{\circ}]$.

The resulting maps are shown in
Fig. \ref{fig:tilt_difazimuths}. Since the data is
effectively sliced in $\phi$, the number of stars at a given $(R,z)$
is lower and as a consequence the spatial bins cover a lower spatial
extent in comparison to Sect. \ref{subsec:axisymmetry}.
A coarse comparison of
the different panels in this figure suggests that the variations with
azimuth are relatively small compared to the global trend that is still apparent
in each panel: the misalignment changes from positive to negative when moving outwards in Galactic radius. 

The most prominent differences are seen for the bins at \mbox{$R\sim4$~kpc} and $z\sim1$~kpc. The 
$\phi=180^\circ$-slice indicates much steeper tilt angles than the $\phi=195^\circ$-slice. The statistical errors on these tilt angles are however large. In fact, most of these bins have consistent 
tilt angles given their error bars.

For a more direct comparison we show in Fig. \ref{fig:exampleradii_difphi}, for specific radii
$R=[6, 8, 10]$~kpc, the tilt angles for the different Galactic azimuths as a function of Galactic height.
Here the different symbols, namely red squares,
green diamonds, and blue crosses correspond to the measurements for
$\phi=[165^\circ,180^\circ,195^\circ]$, respectively. The black
starred symbols show the measurements from all stars at the given $R$ and
$z$ and irrespective of azimuth (as in
Sect. \ref{subsec:axisymmetry}). At $R=10$~kpc the tilt angles for the different azimuths are less consistent with spherical alignment than those at at $R=6$~kpc, especially at positive Galactic heights.

Even though some bins reveal slight differences in the tilt
angles when varying Galactic azimuth, the overall qualitative trends are similar to the case in which we projected all stars onto the $(R,z)$-plane, thus justifying the approach used in
Sect. \ref{subsec:axisymmetry}. These results also suggest that the
degree of non-axisymmetry, in terms of the tilt angles, is modest over the azimuthal range explored.

\begin{figure*}
  \centering
  \includegraphics[width=0.313\textwidth]{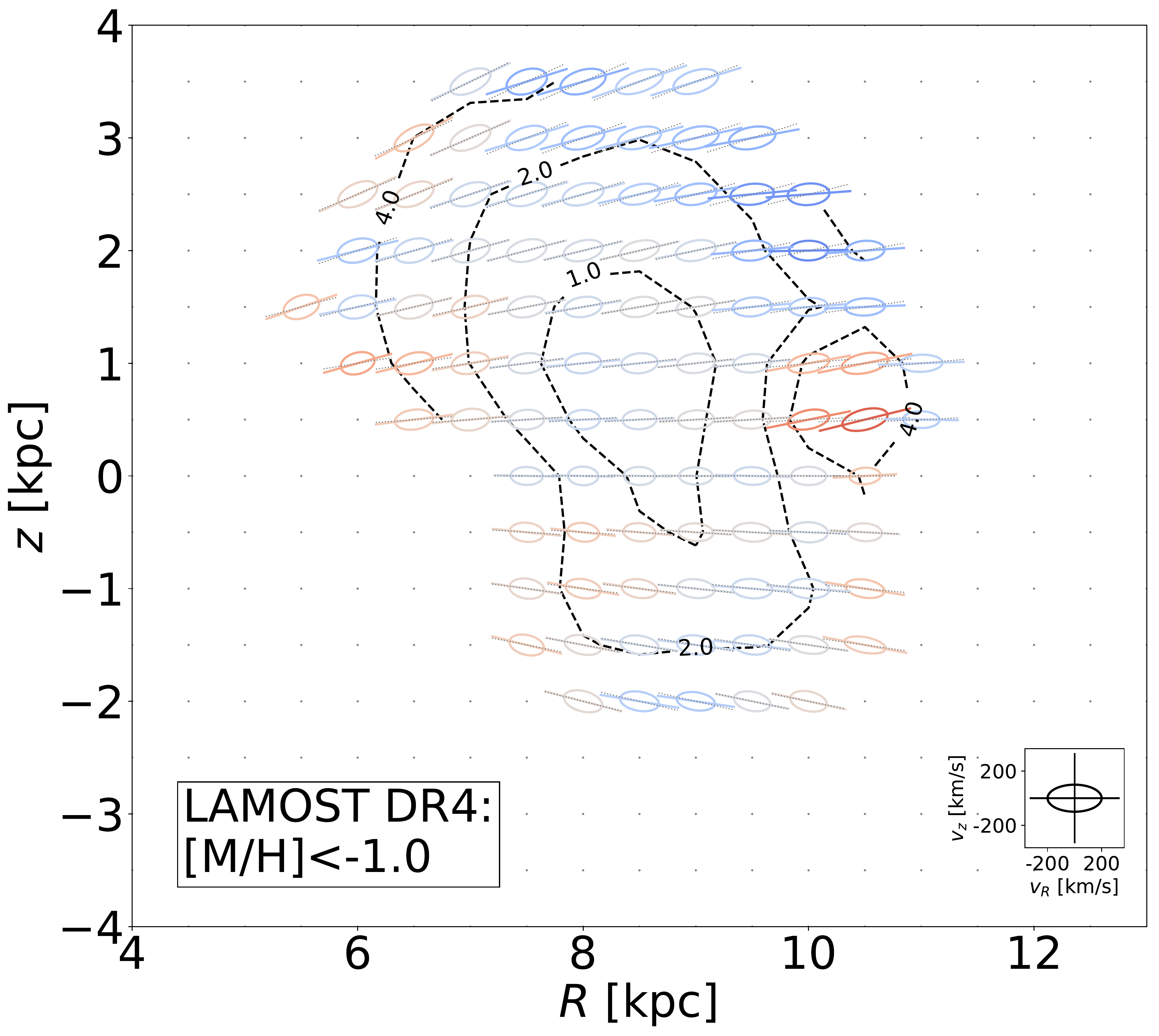}%
  \includegraphics[width=0.315\textwidth]{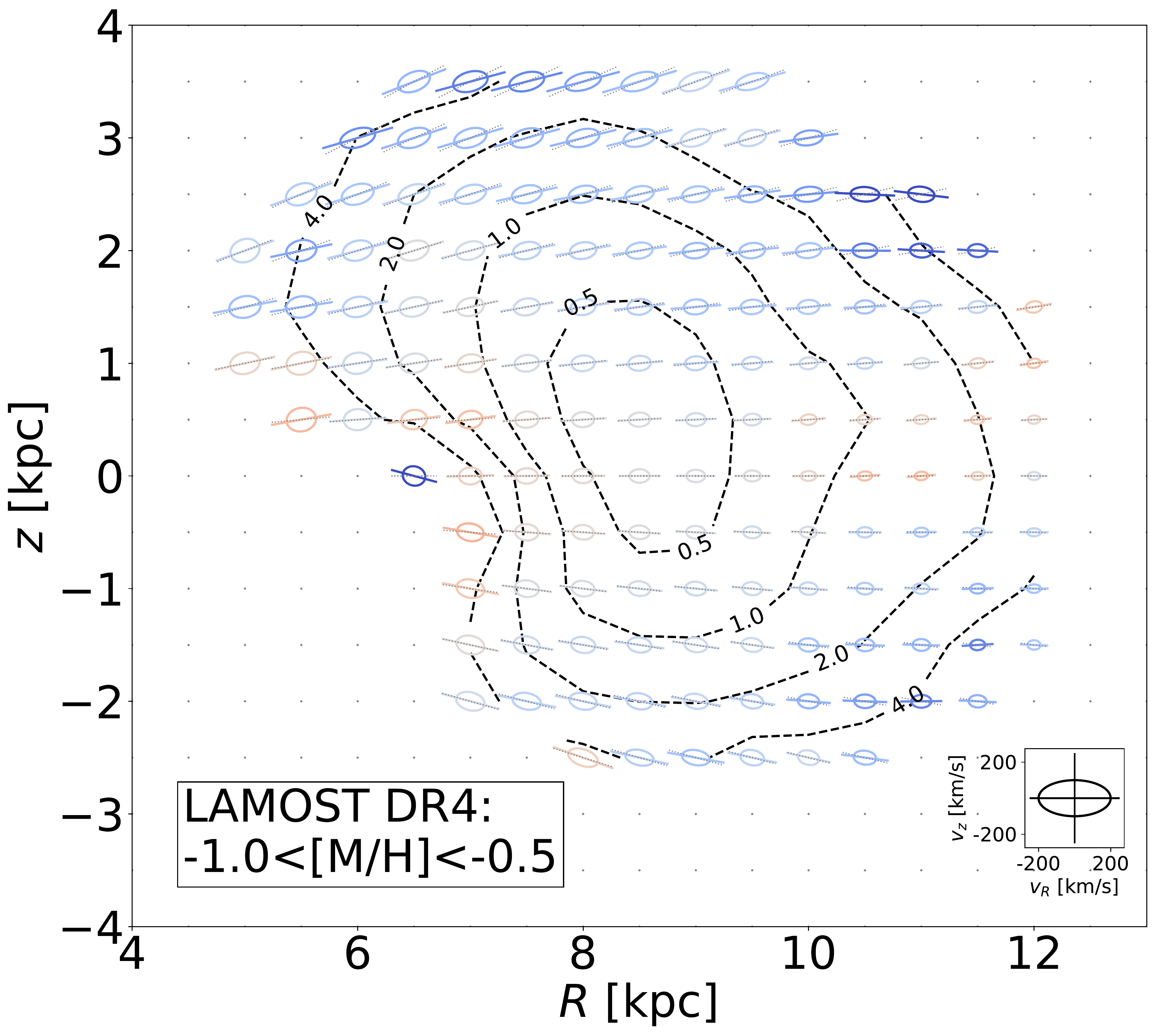}%
  \includegraphics[width=0.372\textwidth]{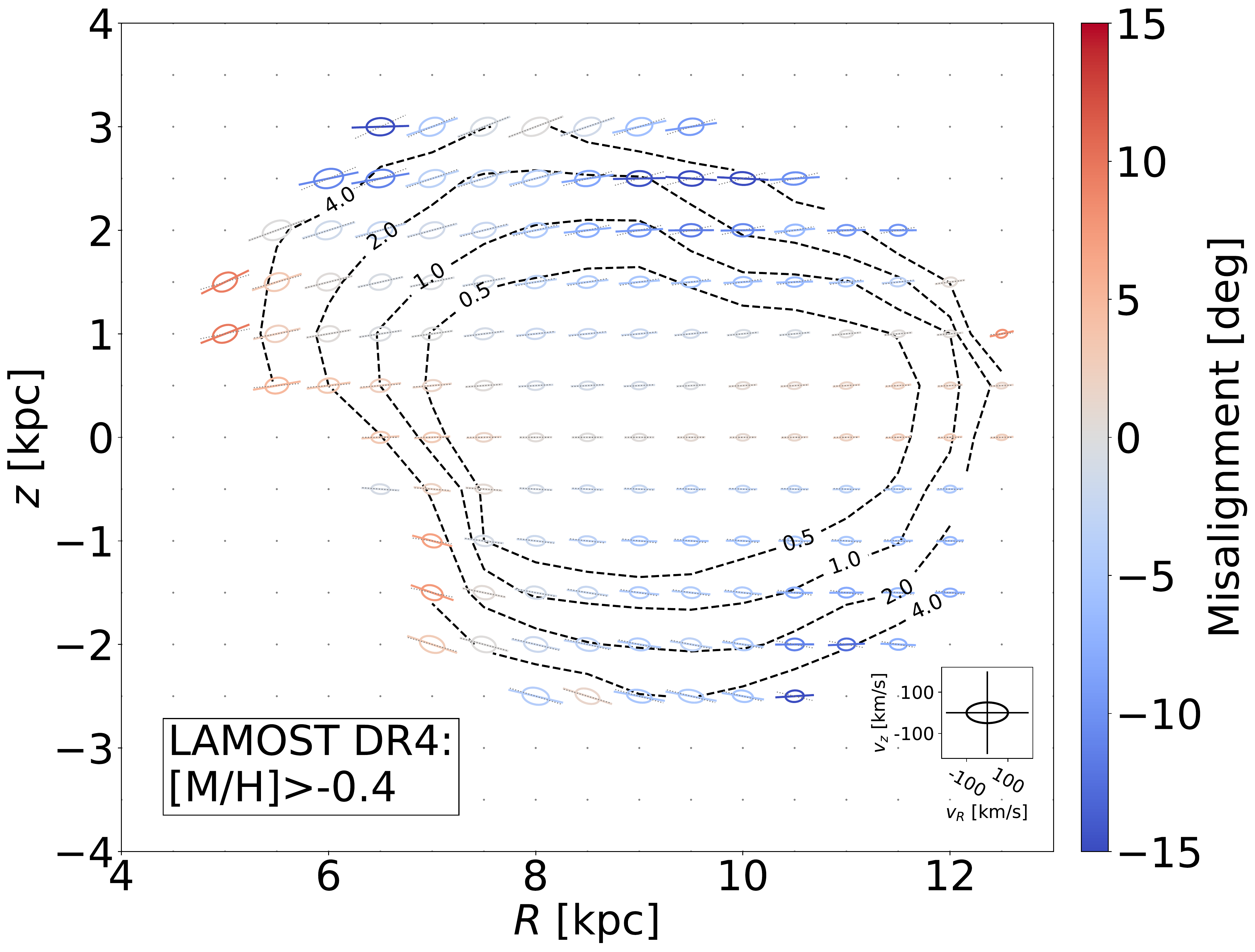}%
  \caption{Velocity ellipses in the meridional plane, as in
    Fig. \ref{fig:misalignment}, but now for the subsamples
    representing halo (left), thick disk (middle) and thin disk
    (right) populations. We note that the scaling of the velocity
    ellipses, indicated by the insets in the bottom right of each
    panel, are different. The colour coding of the ellipses represents
    the misalignment with respect to spherical alignment and is the
    same as in Fig.~\ref{fig:misalignment}. There is no strong evidence
    that the tilt angles of the different populations behave differently.}
  \label{fig:tilt_stellarpops} 
\end{figure*}%

\subsection{Variations with stellar populations}
\label{subsec:stellarpopulations}

In this section we explore whether different populations of stars
follow similar trends in tilt angle. To this end we have cross matched
the full {\it Gaia} DR2 catalogue with three spectroscopic datasets: the
Large Sky Area Multi-Object Fiber Spectroscopic Telescope
\citep[LAMOST DR4, ][]{Cuietal2012}, RAVE DR5, and APOGEE DR14.  If a
star has radial velocity measurements from more than one survey, we
take the measurement with the smallest quoted error. As for the
spectroscopic sample delivered as part of {\it Gaia} DR2
\citep{GaiaDR2_catalogvalidation_Arenouetal2018}, we only consider
stars whose radial velocity errors have been estimated to be smaller
than $20$~km/s.  By adding radial velocities from these other surveys
the number of stars with full phase-space information is increased by
over $30\%$.

To explore dependences on populations, we only use metallicities from
LAMOST DR4 since this survey probes a much larger region than either
RAVE or APOGEE. We refrain from merging the metallicity information
from the different surveys to avoid possible offsets between
metallicity scales. Finally, only stars with metallicity uncertainties
up to $0.2$~dex are considered in our analysis.

A downside of extending our sample is that Bayesian distances are
missing for the newly added stars to our sample. Since the
purpose of this section is to inspect variations between different
populations, we here approximate the distances to the stars by
$\hat{d} = 1/\hat{\varpi}$, where 
\begin{equation}
\label{{eq:updateparallaxes}}
\hat{\varpi} = \varpi + 0.029 \mathrm{\, mas}, \,\,\, {\rm and} \,\,\, 
\hat{\epsilon}_\varpi = \sqrt{\epsilon_\varpi^2 + 0.043^2} \, .
\end{equation}
For the following analysis, we select those stars with at most 20\%
relative distance errors, that is
$\hat{\varpi}/\hat{\epsilon}_\varpi>5$, and $\hat{d}<5$~kpc. We
proceed to classify the stars according to a halo population as those with 
$\mathrm{[M/H]} < -1.0$~dex, a thick disk population for
$-1.0 < \mathrm{[M/H]} < -0.5$~dex, and a thin disk population for
$\mathrm{[M/H]} > -0.4$~dex.  With these criteria, our sample contains
$\sim23,000$ halo stars, $\sim260,000$ thick disk stars, and
$\sim2$~million thin disk stars.

Fig. \ref{fig:tilt_stellarpops} shows the velocity ellipsoids and tilt
angles as a function of position in the meridional plane for the halo
(left), thick disk (middle), and thin disk (right) subsamples. The different spatial coverage of the subsets reflect
differences in the number of stars (recall that to reliably measure a
tilt angle we require at least $100$ stars in a spatial bin). In addition the ellipses for the halo population are much larger compared to those of the thick and thin disks. In fact, we have had to use different
scales for the panels: the insets in the bottom right of each panel
show ellipses whose semi-major and semi-minor axes correspond to
dispersions of $\sigma(v_R)=200$~km/s and $\sigma(v_z)=100$~km/s for
the halo and thick disk populations, and to $\sigma(v_R)=100$~km/s and
$\sigma(v_z)=50$~km/s for the thin disk. 

\begin{figure}
  \centering
  \includegraphics[width=0.4\textwidth]{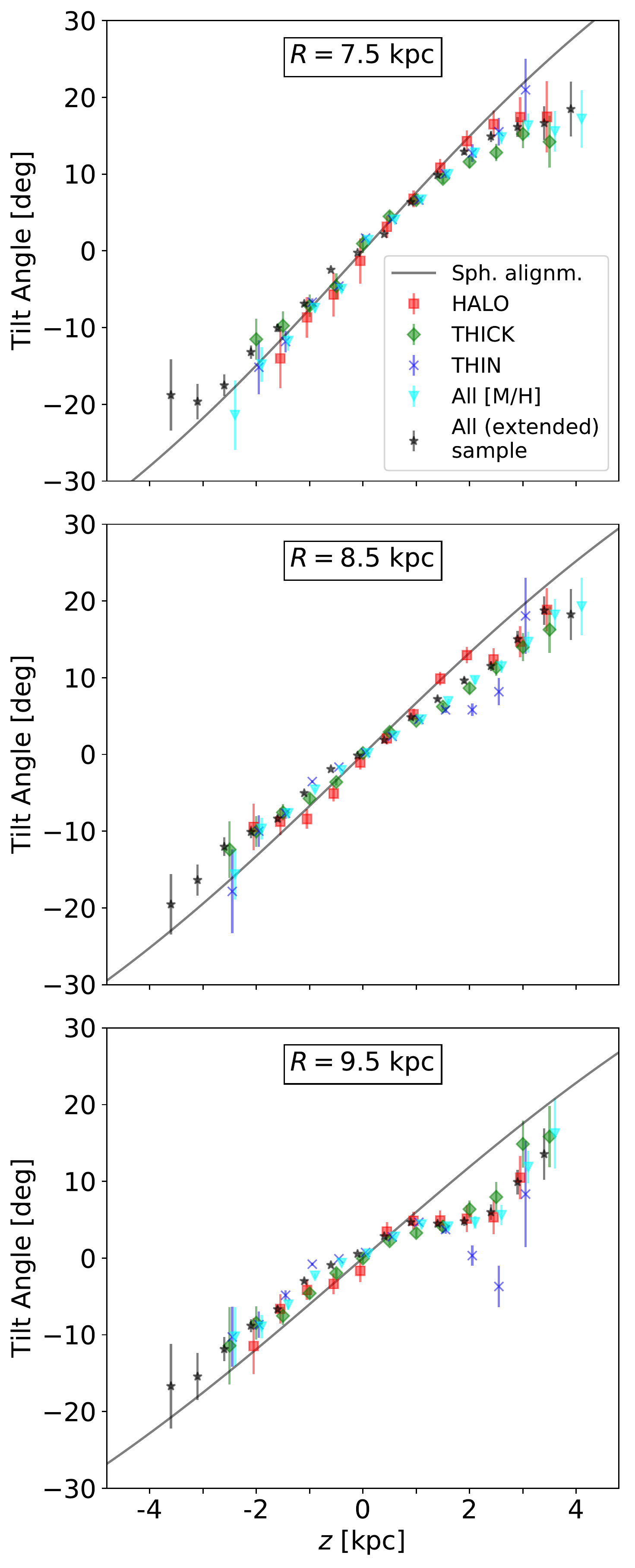}
  \caption{Tilt angles as a function of Galactic height for different
    populations of stars. We show the trends with $z$ for $R=7.5$~kpc (top),
    $R=8.5$~kpc (middle) and $R=9.5$~kpc (bottom). The red squares,
    green diamonds, and blue crosses show the results for the halo,
    thick, and thin disk population described in
    Sect. \ref{subsec:stellarpopulations}, respectively. The light
    blue triangles correspond to all LAMOST stars with metallicity
    information with uncertainties smaller than $0.2$~dex, while the
    black stars are for all stars in the extended sample regardless of
    whether or not they have metallicity information. The solid black
    line shows the trend that would correspond to spherical
    alignment.}
  \label{fig:exampleradii_populations} 
\end{figure}

As in previous sections, the colours in
Fig. \ref{fig:tilt_stellarpops} represent the misalignment of the tilt
angles with respect to spherical alignment. The same
trends as found earlier are visible for the populations
independently: at $R\lesssim 7$~kpc the alignment is closer to
spherical, while outwards from $R\sim9$~kpc the misalignment becomes negative, which means that the tilt angles become shallower. 
This can be seen more easily when comparing the tilt angles derived for each population at specific
radii, as shown in Fig.~\ref{fig:exampleradii_populations}. 

There are also some differences seen. For example, at
\mbox{$R=8.5$~kpc}, the halo sample seems to be more consistent with
spherical alignment than both disk samples.  For $R=9.5$~kpc, however,
the differences between the populations are minor, except for the
flatter thin disk tilt angles at $z\sim2.5$~kpc. Therefore we may
conclude that the results shown in Sect. \ref{subsec:axisymmetry} are
not strongly dependent on the different populations present throughout
the volume probed by our dataset.

\subsection{Quantifying the degree of spherical alignment}
\label{subsec:quantifying}

Because the trends seen in the tilt angles are not strongly dependent on
Galactic azimuth nor on stellar population, we here aim to provide a
simple description of their variation with radius $R$ and height
$z$ as found in Sect. \ref{subsec:axisymmetry}.
Since we infer near spherical alignment for $R \sim 6$~kpc, we
consider expanding $\alpha$ around a point $(R_0,z_0)$: 
\begin{equation}
\begin{split}
\alpha(R,z) = \alpha(R_0,z_0) & + a_1\, (R-R_0) + a_2 \, (z-z_0) \\
                              & + a_3\, (R-R_0)(z-z_0) \\
                              & + a_4 \, (R-R_0)^2 + a_5 \, (z-z_0)^2 + ... \,\,\, ,
\end{split}
\end{equation}
where $a_i$ are constants and both $R$ and $z$ in kpc\footnote{We
  prefer to quantify the deviation from spherical symmetry directly on
  the spherical tilt angle $\alpha$ than to use the purely geometric
  parametrisation by \citet{Binneyetal2014} of the cylindrical tilt angle 
  $\gamma\prime = a_0 \arctan(z/R) = a_0 (\pi/2 - \theta)$ where $\theta$ 
  indicates the spatial location of the bin (see also
  Eq.~\ref{eq:tilt-relations}). Although $a_0 = 1$ implies spherical alignment and $a_0 = 0$ cylindrical alignment, it is not
  intuitively clear what the quantitive meaning of other $a_0$ values is.}. By definition $\alpha(R_0,z_0)=0^{\circ}$.  We further
set $z_0=0$~kpc (i.e. the symmetry plane of $\alpha$ is set to be the
Galactic midplane).  Moreover, $a_1=a_4=0$, since for most realistic
models the tilt angle does not vary at the midplane. By symmetry
arguments the coefficients of all even powers of $z$ (including $a_5$)
must be zero, since $\alpha$ is expected to be either antisymmetric
with respect to the midplane or zero. Since we have found that at
$R \sim 6$~kpc the tilt angles are consistent with spherical
alignment for all $z$ probed (see left panels of
Fig.~\ref{fig:exampleradii}), we additionally set $a_2=0$ such that at
$R=R_0$: $\alpha(R_0,z)=0^\circ$. With these choices:
\begin{equation}
\label{eq:tiltmodel}
\alpha(R,z) \approx a_3\, (R-R_0)z \, .
\end{equation}
We thus fit this functional form to the data to derive values for
$R_0$ and $a_3$ such that the $\chi^2$-statistic defined as:
\begin{equation}
    \chi^2 = \sum_{j=1}^{N_\mathrm{bins}} \left( \frac{\alpha(R_j, z_j)_\mathrm{model} - \alpha(R_j, z_j)_\mathrm{obs}} {\epsilon[\alpha(R_j, z_j)_\mathrm{obs}]} \right)^2 \, .
\end{equation}
is minimised. Here $j$ runs over the number of bins $N_\mathrm{bins}$
where a measurement is made, in other words where $N>100$ stars. 

\begin{figure*}[t]
  \centering
  \includegraphics[width=0.33\textwidth]{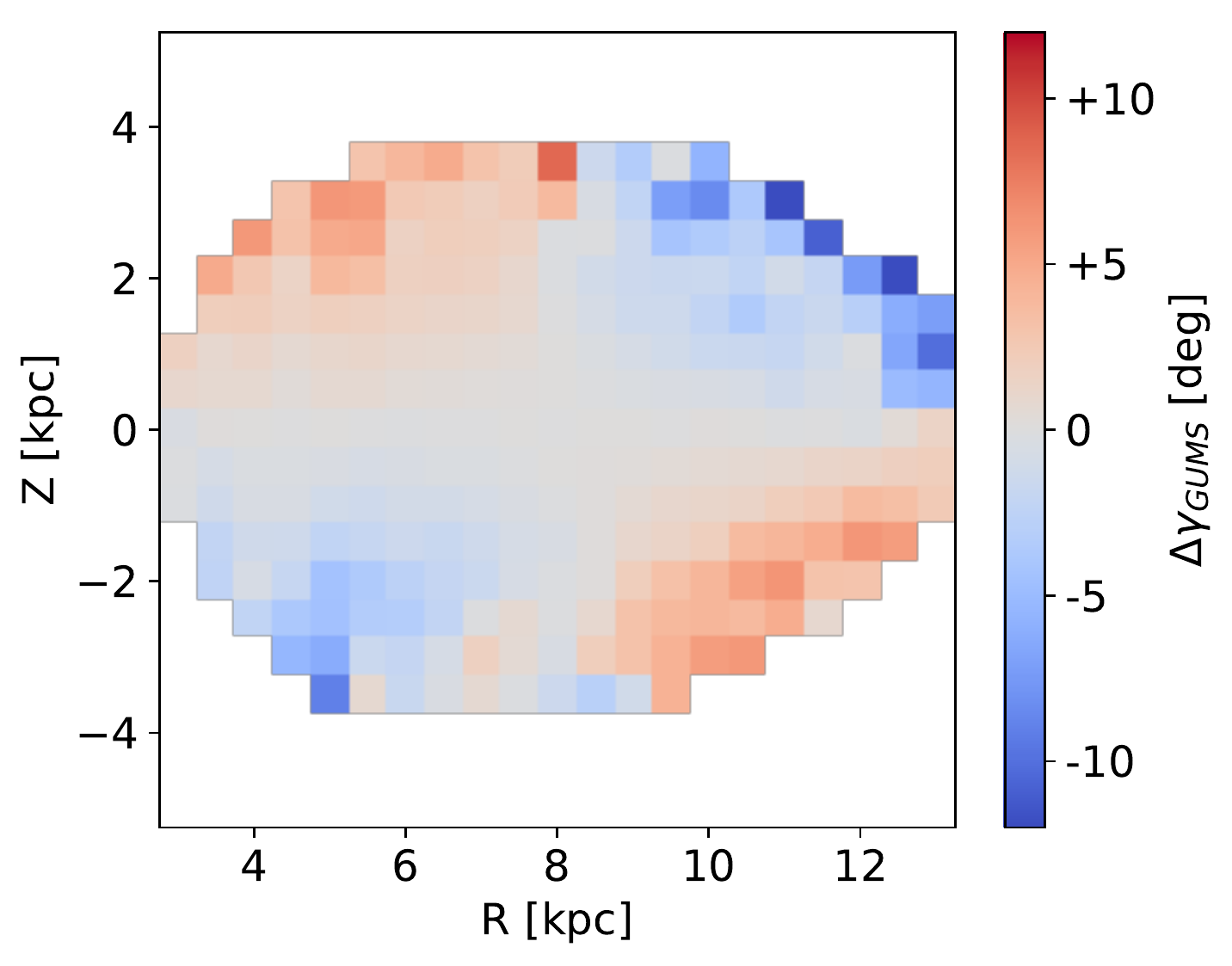}
  \includegraphics[width=0.32\textwidth]{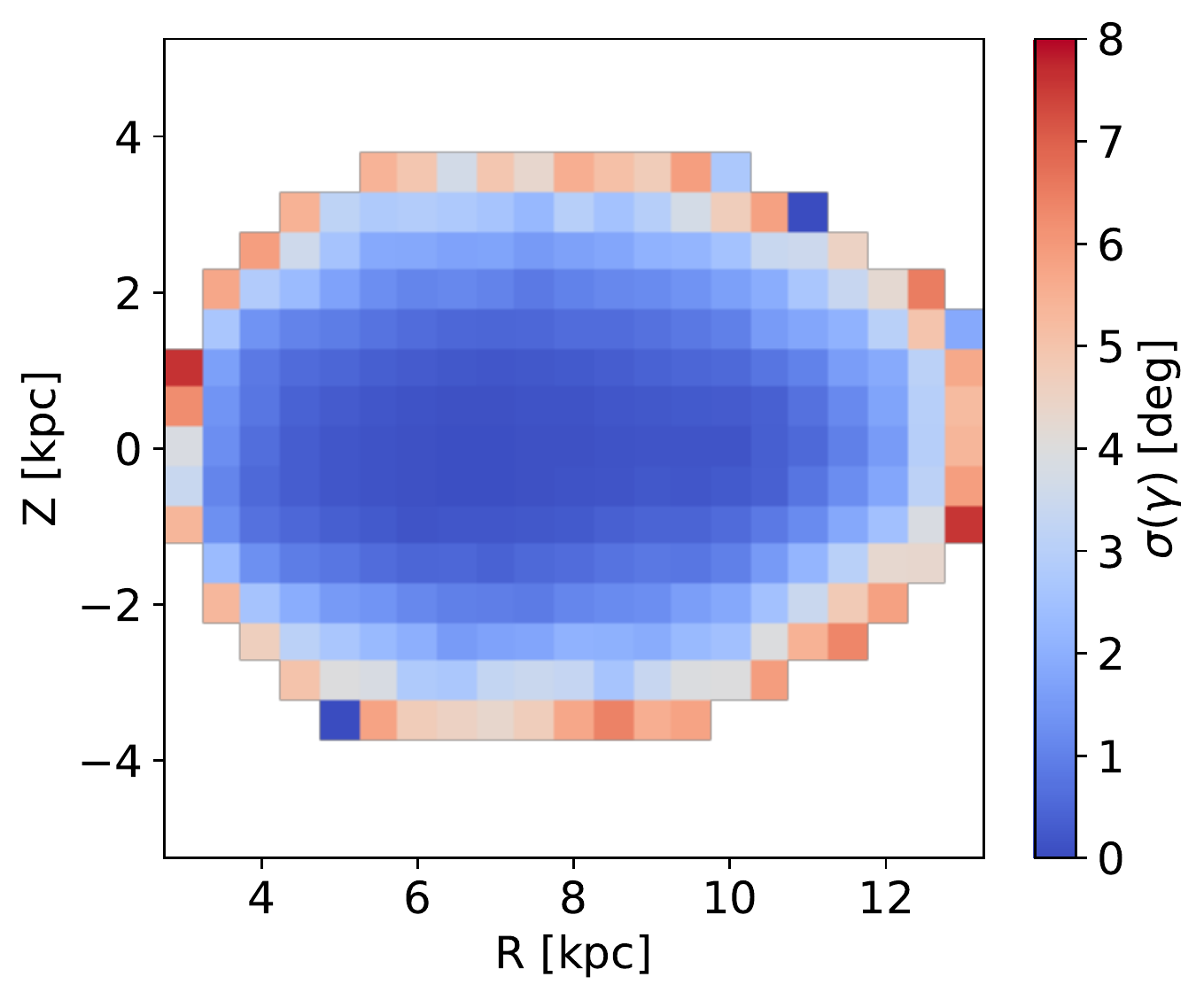}
  \includegraphics[width=0.33\textwidth]{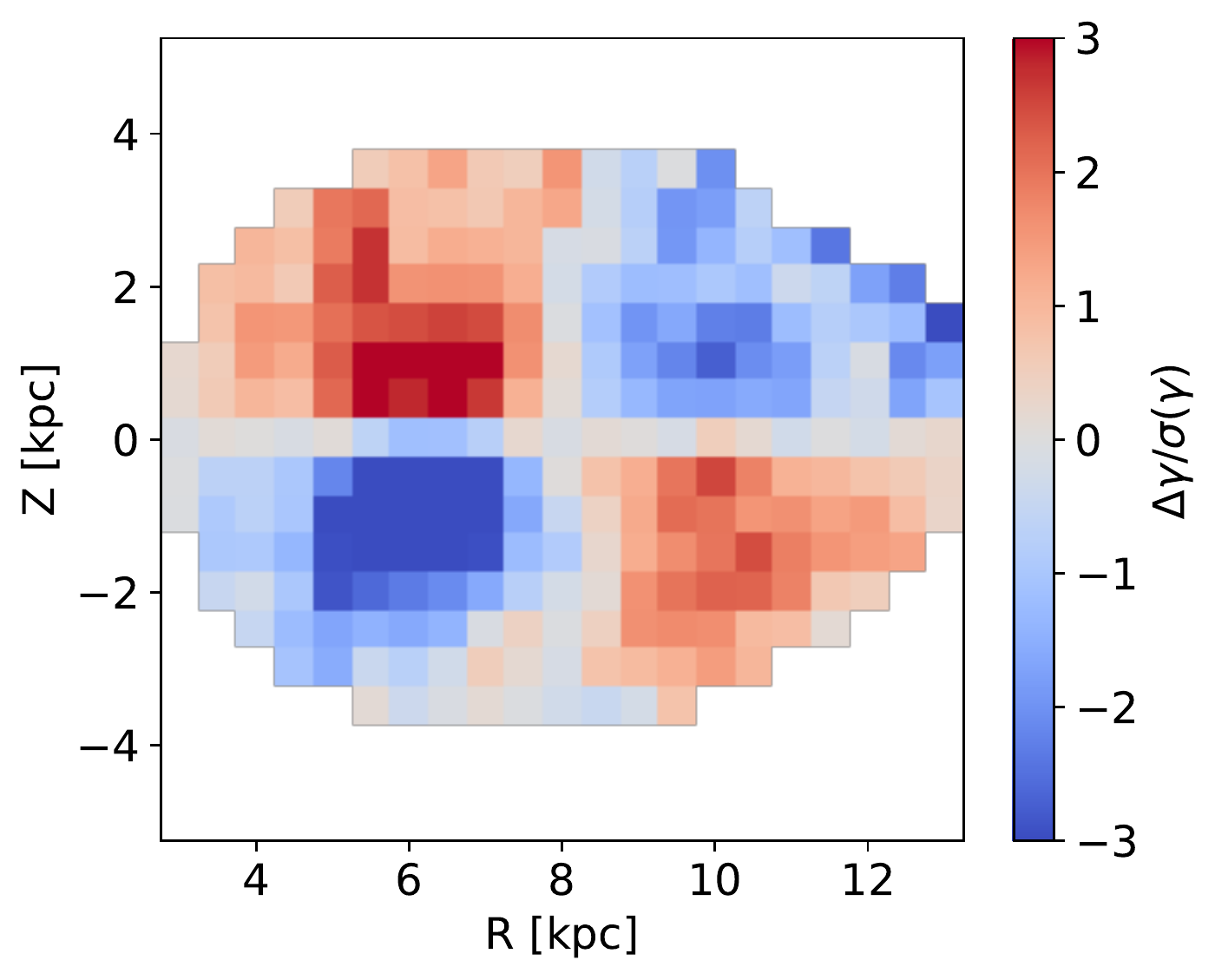}
  \caption{Left: Differences in tilt angles,
    $\Delta\gamma_{\textrm{GUMS}}$, between error convolved
    realisations (taking into account random and systematic parallax
    errors) and the error-free GUMS catalogue.  Centre: Standard
    deviation of the tilt angles over all realisations.  Right:
    Division of the differences by the corresponding standard
    deviation. At distances at around $2$~kpc the changes are
    significant with respect to the scatter present between realisations.}
  \label{fig:GUMSparallaxtestcomplete} 
\end{figure*}%

For most bins at $|z| \leq 2.0$~kpc and $5 \leq R \leq 12$~kpc the
inferred statistical errors on the tilt angles are very small
(e.g. see the dashed contours in Fig.~\ref{fig:misalignment}). In that
case systematic errors need to be considered. One such source of
systematic errors are substructures. We performed tests to estimate
the effect of substructures in velocity space on the tilt angle. To
this end we inserted $N_\textrm{sub}=1$, $4$, $9$, $16$, $25$, or $36$
substructures on smooth non-tilted velocity distributions with
velocity dispersions of $20$~km/s and $35$~km/s in $v_z$ and $v_R$
(i.e. values representative of the thin disk near $R \sim R_\odot$),
respectively. Each substructure was assigned a random number of stars
such that the total fraction of stars in substructures is
$f_\textrm{sub}=5\%$, $10\%$, $15\%$, or $20\%$. We randomly assigned
velocity dispersions to the substructures, drawn uniformly from $1$~km/s to
$5$~km/s in both directions. For each combination of
$(N_\textrm{sub}$, $f_\textrm{sub})$ we considered $100$ realisations.
The median (absolute) tilt angle found from these experiments is $\sim1$ degree,
implying that this value is representative of the error introduced by
neglecting the presence of substructures in a velocity distribution. 
This result is independent of the total number of stars $N$ for $N\gtrsim10,000$ (a value that is representative of the number of stars in the bins with $\epsilon[\alpha(R_j, z_j)]<1^\circ$). 
Thus, when minimising the $\chi^2$ we consider a floor for the statistical error $\epsilon[\alpha(R_j, z_j)]$
in each bin of $1^\circ$. 

We fit to find $R_0=(6.16 \pm 0.16)$~kpc and
$a_3=(0.72 \pm 0.04)^\circ/{\mathrm{kpc}}^2$ resulting in a reduced
$\chi^2$ of $1.65$. The cyan line in Fig. \ref{fig:exampleradii} shows
the tilt angles predicted by this fit, which reproduces relatively
well the trends observed in the data. The model goes through the
$1\sigma$-error bars for approximately $60\%$ of all spatial bins, while
for $98\%$ of bins the model matches the data within 3$\times$ the
estimated uncertainty. This indicates that our simple model provides a fair
description of the behaviour of the tilt of the velocity ellipsoid
across the Galactic volume probed by our dataset.

The fact that the total reduced $\chi^2$-value is greater than unity indicates that the tilt angles for some bins are not fitted very well by the model. For example at $R\sim10$~kpc the tilt angles as inferred from the data are asymmetric with respect to the $z=0$ plane: at $z > 0$~kpc they more
or less attain a constant value of $\sim 2.0^\circ$, whereas below the
midplane the tilt angles become steeper with $z$ (e.g. $-15^\circ$ at
$z=-3.0$~kpc). The fits at such radii are therefore relatively
poor. For the bins between $R=11$~kpc and $R=12$~kpc, we notice that the observed tilt angles seem to have a small
positive offset from zero near $z=0$. These offsets are small (of order $2$
degrees), although they do affect the goodness of fit measure.




\section{Discussion}
\label{sec:discussion}



\subsection{The impact of (parallax) measurement errors on the recovered tilt angles.}
\label{subsec:parallaxoffset}

\cite{GaiaDR2_summary_Brownetal2018} have reported the presence of a
systematic error on the {\it Gaia} DR2 parallaxes in the form of a
zero-point offset of a few 10 of $\mu$as (in the sense that
\textit{Gaia} parallaxes are too small) and whose exact amplitude
depends on location on the sky. Such systematic zero-point offset
affects the tangential velocities of the stars, which are determined
from both distances and proper motions.  The overall systematic parallax
offset in \textit{Gaia} DR2 was determined using distant quasars by
\citet{GaiaDR2_astrometricsolution_Lindegren2018} to be approximately
$-29\mu$as, with a large RMS of $\sim 43
\mu$as. \citet{GaiaDR2_catalogvalidation_Arenouetal2018} using
different samples of objects (RR Lyrae stars, Magellanic Clouds, open
clusters, dwarf spheroidal galaxies, etc.) report important variations
in the zero-point offsets, highlighting the complexity of the
offset. Nonetheless all values are consistent given the large
estimated RMS.

Around the time this paper was submitted,
\citet{Schonrichetal2019_distancesGaiaDR2_rvssample} reported a new
estimate of the parallax zero-point offset based on the distance
estimation method used in \citet{Schonrich&Aumer2017_distancesTGAS}
\citep[also see][]{Schonrichetal2012_detection_of_distance_errors}. These authors argue for a much larger
zero-point for the parallaxes in the RVS subset of \textit{Gaia} DR2,
namely of magnitude $-54 \pm 6
\mu$as. \citet{Zinnetal2018_GaiaDR2_offset_RGB_and_RCstars} and \citet{Khanetal2019_GaiaDR2_offset_RGB_and_RC_stars}
applied asteroseismology to determine distances to Red Giant Branch
(RGB) and Red Clump (RC) stars with \textit{Gaia} $G$-band magnitudes
similar to those present in the RVS subset of \textit{Gaia} DR2 and
determined an offset close to $-50 \mu$as, while \citet{Sahlholdt2018},
using asteroseismology information on dwarfs, report that the offset
could be $\sim -35 \pm 16 \mu$as. More recently \citet{Hall2019},
using RC stars with asteroseismology, estimate the mean offset to be
$-41 \pm 10 \mu$as. These comparisons suggest that the offset could
well be larger for the brighter stars of the {\it Gaia} RVS sample but
that its amplitude is quite uncertain.

\subsubsection{Quantification of the impact of a zero-point offset}

We first quantify how the tilt angles are affected if parallaxes are
underestimated.  For illustration purposes, we estimate the impact on
the recovered tilt angles induced by a systematic error (with mean
$-29 \mu$as) while also including the effects of random
errors\footnote{In Appendix \ref{App:Appendix_parallaxoffset} we
  analytically compute how the $v_R$- and $v_Z$-velocities (and thus
  their moments and tilt angles) are affected by the parallax
  zero-point offset alone.}. Their effect is examined by using the
\textit{Gaia} Universe Model Snapshot \citep[GUMS,
][]{Robinetal2012_GUMS}, which is based on the Besan\c{c}on Galaxy
Model \citep{Robinetal2003_Besancon}. 

We mimic the \textit{Gaia} DR2 subsample with full phase-space
information, by selecting stars in GUMS that have $G<13$ mag, as this
is roughly the magnitude limit for radial velocities in
\textit{Gaia}'s current data release. We generate $100$ data
realisations by convolving the (error-free) GUMS sample with a
Gaussian with \textit{Gaia} DR2-like random and systematic errors for
the parallaxes \citep{GaiaDR2_astrometricsolution_Lindegren2018}. The
systematic parallax offsets for the stars are drawn from a Gaussian
with mean $-29 \mu$as and standard deviation of $30 \mu$as\footnote{as
  estimated in \citet{GaiaDR2_kinematicsofsatellites_Helmietal2018}.}.
To obtain a distance estimate we invert the parallaxes and consider
only those stars that satisfy $\varpi/\epsilon(\varpi)>5$ and
$\varpi \gtrsim 200 \mu$as. Here $\varpi$ is the observed parallax and
$\epsilon(\varpi)$ the random parallax error and thus the same quality
criteria are applied as to the real data (see Sect. \ref{sec:data}).

\begin{figure*}[t]
  \centering
   \includegraphics[width=1.0\textwidth]{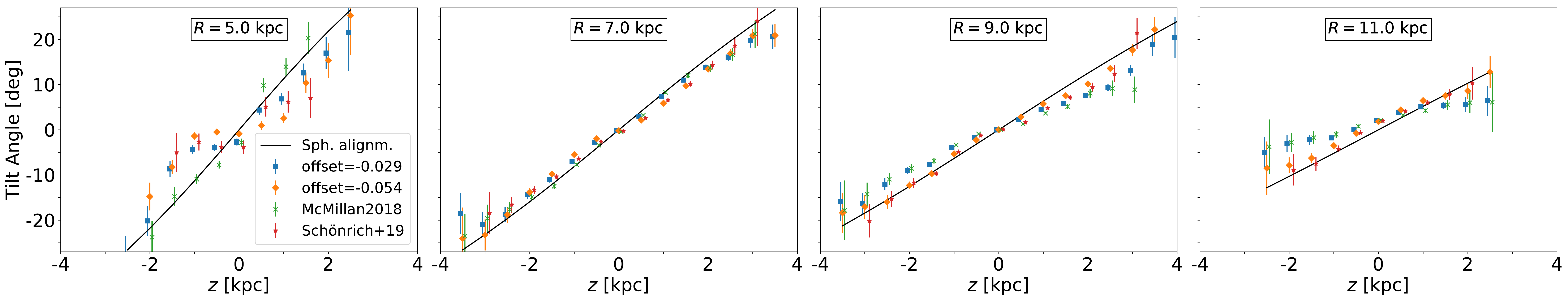}
   \caption{Tilt angles as a function of Galactic height for different
     positions across the Galaxy. We show the trends with $z$ for
     $R=[5,7,9,11]$~kpc for different distance estimates for the
     stars. The blue squares and orange diamonds use distances based on
     inverting the parallaxes after correcting the parallaxes for an
     offset of $-29 \mu$as and $-54 \mu$as, respectively. The green crosses
     and red starred symbols use Bayesian distances from
     \citet{McMillan2018_simpleGaiaDR2distances_RVSsubset} and \citet{Schonrichetal2019_distancesGaiaDR2_rvssample}, respectively. The solid
     black line shows the trend that would correspond to spherical
     alignment.}
  \label{fig:exampleradii_compareoffsets} 
\end{figure*}%

For each spatial bin the median (over all realisations) of the
distribution of tilt angles is compared to the tilt angles from the
error-free model, on the meridional plane.  The error-free GUMS model
has close to cylindrically aligned velocity ellipses
($\gamma_{\textrm{GUMS}} \sim 0^\circ$). The impact of the random and
systematic parallax uncertainties on the tilt angles depends on
location as can be seen in the left panel of
Fig.~\ref{fig:GUMSparallaxtestcomplete}.  At $R\lesssim7$~kpc the
orientations of the velocity ellipses change towards the direction of
spherical alignment ($\Delta \gamma_{\textrm{GUMS}} > 0$ for $z>0$ and
$\Delta \gamma_{\textrm{GUMS}} < 0$ for $z<0$), while for
$R \gtrsim 9$~kpc the change is in the opposite sense.

The middle panel of Fig.~\ref{fig:GUMSparallaxtestcomplete} shows the
spread in tilt angles over all realisations, and reveals that the
errors result in a spread with a typical amplitude of $\lesssim 4^\circ$, except
for the outermost bins, where it can be twice as large, and hence
comparable to $\Delta \gamma_{\textrm{GUMS}}$. The right panel shows
at which locations the median change in tilt angles, caused by
parallax errors, is larger than the RMS from realisation to
realisation. For bins located at distances of $\sim 2$~kpc a change in
tilt angle due to parallax errors is thus likely to occur in a
preferential direction, with the amplitude of this change varying from
realisation to realisation.

These findings imply that, if parallaxes are underestimated, the tilt
angles inferred may appear steeper than they really are in the inner
Galaxy, while the opposite happens in the outer Galaxy, thus the tilt
angles become shallower there. If we take the results from GUMS at face
value, $|\Delta \gamma_{\textrm{GUMS}}| \approx 6^\circ$ at
$(R,|z|) \sim (5,3)$~kpc, which means that an unaccounted for
zero-point offset of magnitude $29 \mu$as in the parallaxes affects the inferred tilt
angles such that they appear steeper by $\sim6^\circ$. This does not
radically change the type of alignment at this location (where
spherical alignment would imply $\gamma \sim 30^\circ$). For
$(R,|z|) \sim (11,2)$~kpc we find that
$|\Delta \gamma_{\textrm{GUMS}}|$ can attain values close to
$5^\circ$, which is of similar amplitude as the misalignment seen in
Fig. \ref{fig:misalignment}.  Although the GUMS tilt angles
intrinsically have $\gamma_{\textrm{GUMS}} \sim 0^\circ$, we find
similar amplitudes for the cases explored in Appendix
\ref{App:Appendix_parallaxoffset}, where we start from both
intrinsically spherically and cylindrically aligned ellipsoids.

In the analysis presented in previous sections, we have effectively
corrected for the parallax offset by using the
\citet{McMillan2018_simpleGaiaDR2distances_RVSsubset} distances. If the assumed parallax
zero-point is too small, the results presented in this section indicate that, especially towards
the outer Galaxy, the zero-point offset could produce tilts that are
less steep than what they are intrinsically. We explore such a larger offset next. 


\subsubsection{A zero-point offset as large as -54$\mu$as}
\label{subsec:schonrich2019}

\citet{Everalletal2019} have derived tilt angles using the
\citet{Schonrichetal2019_distancesGaiaDR2_rvssample} Bayesian distance
estimates (with parallax zero-point of $-54 \mu$as). These authors
showed that the tilt angles appear to be much more consistent
with spherical alignment when using those distances.

Since the method used in
\citet{Schonrichetal2019_distancesGaiaDR2_rvssample} assumes spherical
alignment, we preferred not to directly use their distances while testing for the effect of a large $-54 \mu$as offset.
Therefore we here also explore 
how the tilt angles change if the parallax offset would be as large as
$-54 \mu$as, by comparing them to the case in which the offset is
$-29 \mu$as. For both cases we take the 
extended sample and invert the parallaxes after correcting for the zero-point offset (as in
Sect. \ref{subsec:stellarpopulations}), such that the changes due to
the differences in parallax offset can be easily compared.

In Fig. \ref{fig:exampleradii_compareoffsets} we show the results. The
blue squares have been calculated after correcting for a parallax
zero-point offset of $-29\mu$as, whereas for the orange diamonds a
value of $-54\mu$as is assumed. For the outer Galaxy ($R=9$~kpc and
$R=11$~kpc) such a larger parallax zero-point can modify the tilt angles
such that they are more consistent with spherical alignment, in agreement with our analysis of the previous
section. 

A direct comparison of the tilt angles obtained using
\citet{McMillan2018_simpleGaiaDR2distances_RVSsubset} Bayesian
distances (who assumes a zero-point of $-29 \mu$as, green crosses) with
the results obtained from inverting the parallaxes after correcting
for a zero-point of $-29 \mu$as (blue squares), shows good agreement
except for $R=5$~kpc. At this location, it would seem as if the choice of the distance
estimator would play a role in the determination of the tilt
angle. The Bayesian distances result in tilt angles that are just
slightly steeper than expected for spherical alignment, while
inverting the parallaxes results in much shallower tilt angles (the
larger the offset assumed the shallower the tilt angles).
On the other hand, comparing the tilt angles obtained using
\citet{Schonrichetal2019_distancesGaiaDR2_rvssample} Bayesian
distances (who find a zero-point of $-54 \mu$as, red starred symbols), with
the results obtained from inverting the parallaxes after correcting
for a zero-point of $-54 \mu$as (orange diamonds), shows rather similar 
trends at $R=5$~kpc. At the other radii shown, these Bayesian distances
also result in tilt angles that are in good agreement with inverting the parallaxes.

The analysis presented in the last two sections shows that the
amplitude of the systematic error in the parallax, in the form of a
zero-point offset, plays a role in the determination of the tilt
angles for the outer Galaxy ($R > 9$~kpc). Since the offset is known
to vary with celestial position, magnitude and colour, it is difficult
at this point to properly correct for it, and this impairs a very
accurate determination of the tilt angle throughout the range of
distances probed. However, recall that the range of zero-point offsets is bracketed by the values
explored (i.e. from $-54\mu$as to $-29\mu$as), so the analysis presented
here gives us a handle on the possible outcomes.


\subsection{Constraints to models of the Milky Way}
\label{subsec:massmodels}

Several models of the Milky Way have been proposed by matching a
variety of constraints
\citep[e.g.][]{McMillan2011,Piffletal2014_constraininghalowithRAVE,
  McMillan2017}. Particularly useful for the interpretation of the
findings reported in this paper are St\"{a}ckel models \citep[e.g.][]{deZeeuw1985, Dejonghe&deZeeuw1988}.  
Axisymmetric models with a potential of St\"{a}ckel form have the property that the
equations of motion are separable in their spheroidal coordinates. 
Therefore the principal axes of the
velocity ellipsoids are always aligned with these coordinates \citep[also see:][]{Eddington1915}. The
foci of such a coordinate system then determine the alignment at each
position. For a composite model to be of a St\"{a}ckel form, the
locations of the foci must be identical for all components.

\begin{figure}[t]
  \centering
  \includegraphics[width=0.5\textwidth]{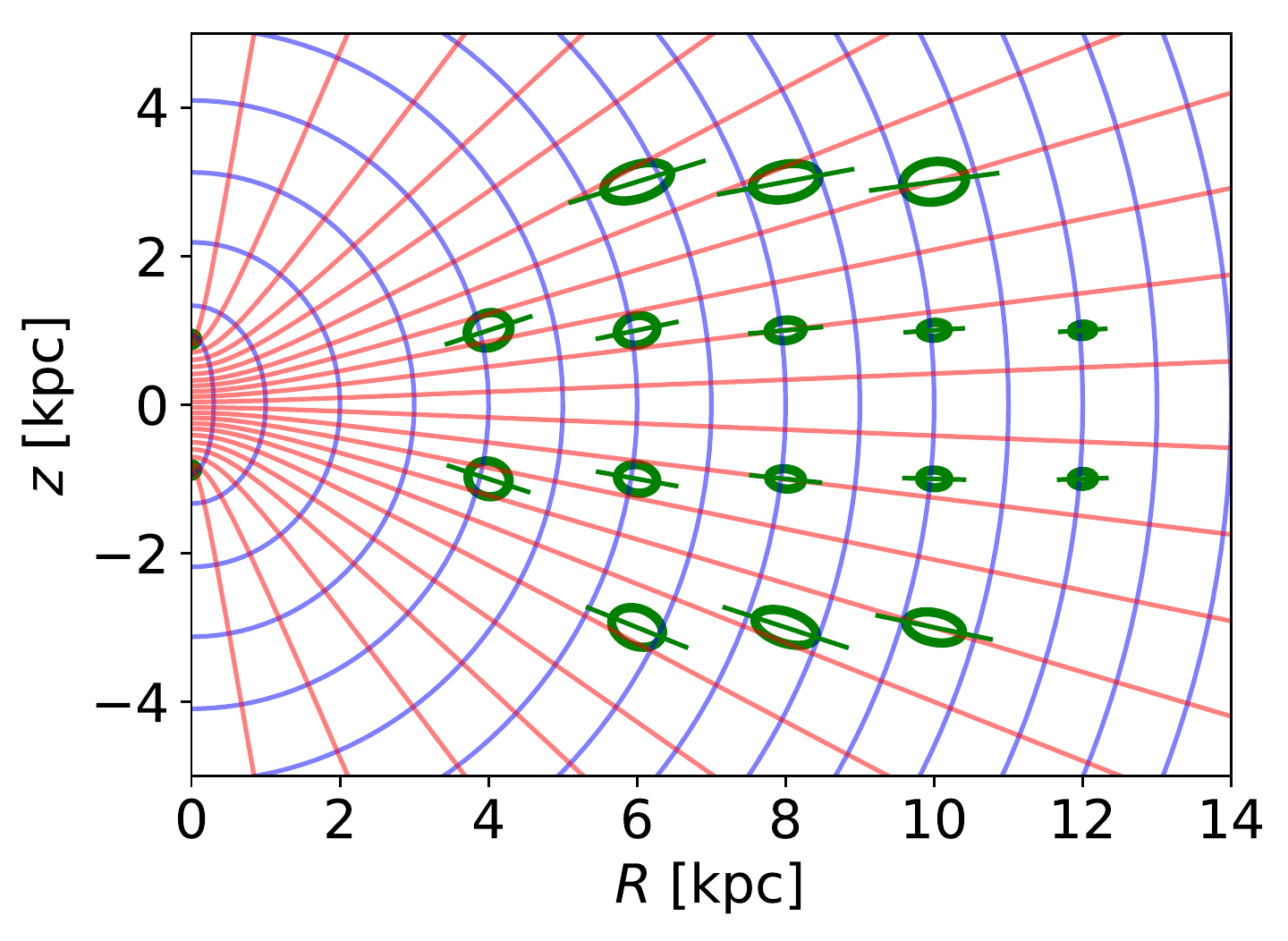}
  \caption{Contours of constant prolate spheroidal coordinates,
    $(\lambda, \nu)$, with foci at $R=0$ and $z=\pm 0.88$~kpc (see
    text). Contours of constant $\lambda$ are shown in blue, contours
    of constant $\nu$ in red. The green ellipses show some of our measured velocity ellipses (Method 1).
    Their orientation does not align with the coordinate contours at $R\gtrsim10$~kpc and $|z|\gtrsim2$~kpc.}
  \label{fig:tilt_Stackel} 
\end{figure}

\citet{Famaey&Dejonghe2003}, for example, have extended the two-St\"{a}ckel
component work of \citet{Batsleer&Dejonghe1994} by adding a third
component, such that the model could allow for a thin and thick disk,
in addition to a halo component.  The authors use constraints such as
the (flat) rotation curve, circular velocity at the position of the
Sun, the Oort constants, and the local total mass density in the disk
to search for a set of consistent parameters for their St\"{a}ckel
models.
Here we take the set of prolate spheroidal coordinates, $(\lambda, \phi, \nu)$, 
from \citet[][mass model II]{Famaey&Dejonghe2003}. 
The foci of this oblate mass model are located at ($R$, $z$) = ($0$,
$\pm 0.88$)~kpc. At $R\sim0$ and $|z| \lesssim 0.88$~kpc such spheroidal
coordinates align with the cylindrical coordinate system (see Fig. \ref{fig:tilt_Stackel}). Outside of
these foci and with increasing distance from the Galactic centre the
spheroidal coordinates approach the spherical coordinate system. 
In general, any (composite) St\"{a}ckel model predicts a change in the
tilt of the velocity ellipse from cylindrical to spherical
alignment. The transition radius depends on the location of the foci.

Since the observed tilt angles at $R\sim4$~kpc already show near
spherical alignment, this implies foci at $|z|\lesssim 4$~kpc. Their
exact position would depend on whether the innermost region of the
Galaxy, not probed by our dataset, is cylindrically aligned or not,
and if so at what distance the transition occurs.  However, the tilt
angles in the outer Galaxy ($9 \lesssim R \lesssim 12$~kpc) derived
using the \citet{McMillan2018_simpleGaiaDR2distances_RVSsubset}
distances are not consistent with St\"{a}ckel models that have foci at
$|z|\lesssim 4$~kpc, and would require a larger focal
distance. We have numerically checked these statements by comparing the predicted tilt
angles of both oblate and prolate St\"{a}ckel models (for a large
range of different focal distances) to the observed tilt angles while
taking into account their errors.  

\begin{figure}
  \centering
  \includegraphics[width=0.5\textwidth]{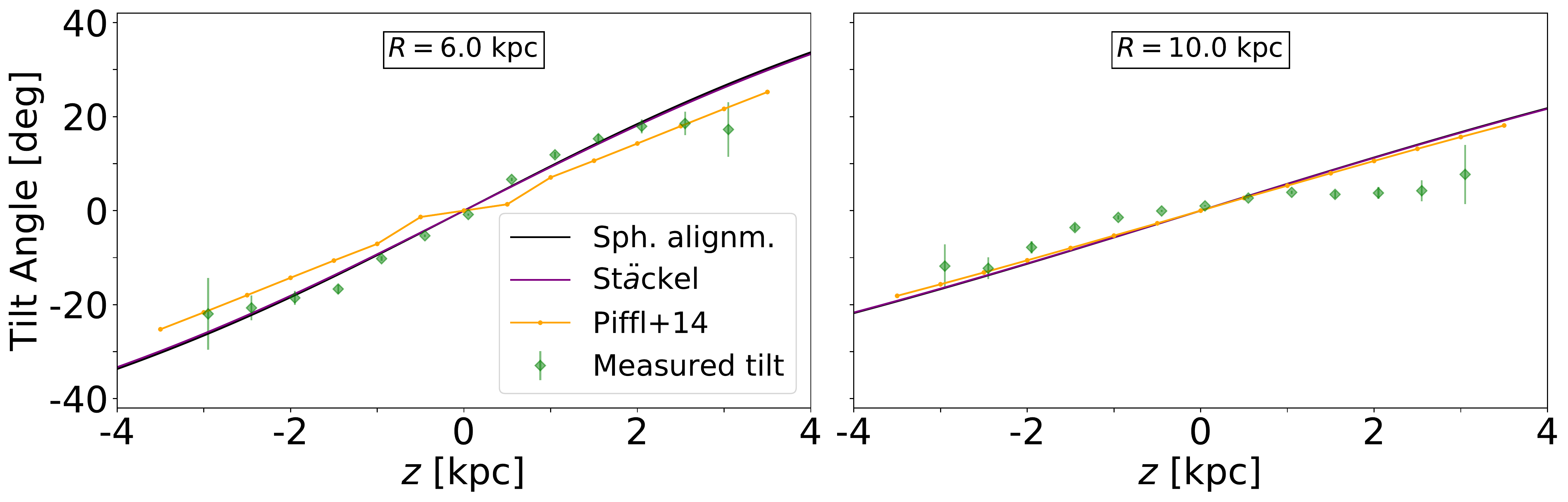}
  \caption{Tilt angles for both the St\"{a}ckel (purple line) and \citet[][orange line]{Piffletal2014_constraininghalowithRAVE} model for radii at $R=4$~kpc and $R=8$~kpc (see text). For comparison we add our measurement as green diamonds (Method 1).} 
  \label{fig:tilt1D_Stackel_Piffl} 
\end{figure}%

There are of course many more models with bulge, disk and halo components, for example spherical bulge, exponential disk, Navarro-Frenk-White \citep[NFW,][]{NFW1996} halo, or \citet{Miyamoto&Nagai1975} models. The separable models are in that sense a subset but have the advantage that for them the tilt of the velocity ellipsoid is dictated by the coordinate system in which the equations of motion (Hamilton-Jacobi equation to be more precise) separate. 

\citet{Piffletal2014_constraininghalowithRAVE} have applied a five
component mass model (gas disk, thin and thick disk, flattened bulge
and dark halo) to RAVE DR4 stars. Using their best-fitting parameters
we computed the relevant velocity moments from the distribution
function for a similar range in $R$ and $z$ as probed in our
dataset. The tilt angles for this model are spherically aligned for
$R\gtrsim7$~kpc and are, as in the separable models discussed above,
changing towards cylindrical alignment with decreasing $R$.

In Fig. \ref{fig:tilt1D_Stackel_Piffl} we show the tilt angles for
both the St\"{a}ckel model (purple line) of
\citet{Famaey&Dejonghe2003} and the
\citet{Piffletal2014_constraininghalowithRAVE} model (orange line),
for radii $R=6$~kpc and $R=10$~kpc. The green diamonds indicate the tilt
angles as found by Method 1. Since this St\"{a}ckel model has
focii at $|z| \lesssim 0.88$, which is very close to the Galactic
centre with respect to the innermost radius probed in our dataset, the
St\"{a}ckel model is almost indistinguishable from spherical alignment
for all positions probed. The \citet{Piffletal2014_constraininghalowithRAVE} model has tilt angles that are
shallower at $R=6$~kpc, but also approaches spherical alignment with
increasing Galactic radius. At $R=10$~kpc, for example, the tilt angles from the
\citet{Piffletal2014_constraininghalowithRAVE} model are seen to nearly coincide with the
expectation for spherical alignment.

We note that if the parallax zero-point is larger than assumed here the
tilt angles do become more consistent with spherical alignment for
large radii (see \ref{subsec:schonrich2019}). This is in line with
predictions for both composite St\"{a}ckel models as well as for the
\citet{Piffletal2014_constraininghalowithRAVE} model. 
In addition, it would be interesting to know whether
the tilt angles become shallower towards the central regions of the
Galaxy (at $R\lesssim4$~kpc). In principal it would then be possible
to solve for the focal distance. However, the effects of both the type
of distance estimator and the assumed parallax zero-point are too
large to make firm statements in this region. Future data releases
will for sure enable to probe regions closer to the Galactic centre
more robustly.


\section{Conclusions}
\label{sec:conclusions}

We have studied the trends in the tilt angle of the velocity
ellipsoids in the meridional plane for a high-quality sample of more than $5$
million stars located across a large portion of the Galaxy, from
$R \sim 4$~kpc to $R \sim 13$~kpc, and reaching a maximum distance from the plane
of $\sim3.5$~kpc. 

We find that the tilt angles are somewhat dependent on the 
offset of the \textit{Gaia} DR2 parallaxes, and that the effects
are particularly important for the outer Galaxy. When using the
\citet{McMillan2018_simpleGaiaDR2distances_RVSsubset} Bayesian
distances, derived assuming an offset of $-29 \mu$as, we find that the
tilt angles are consistent with (near) spherical alignment at
$R\lesssim 7$~kpc for all heights probed ($|z|\lesssim3$~kpc). Beyond
$R \gtrsim 9$~kpc the tilt angles clearly become more shallower than
expected for spherical alignment. These trends remain when the stars
are separated into `populations' according to their metallicity (as
given by LAMOST DR4). We provide a simple analytic function for the
tilt angle in spherical coordinates
$\alpha(R,z)/[{\rm deg}] \approx 0.72 (R-6.16)z$, that fits well the trend
observed as a function of Galactic radius and height, after projecting
the stars onto the $(R,z)$-plane.

We find that if the amplitude of the zero-point offset in the
parallax is underestimated, the angles tend to appear shallower
than they intrinsically are in the outer Galaxy (i.e. changing into
the direction of cylindrical alignment if the ellipsoid is
intrinsically spherically aligned). We quantify the impact on
the tilt angles when assuming a parallax zero-point as large as
$-54 \mu$as, as estimated in \citet[][]{Schonrichetal2019_distancesGaiaDR2_rvssample} \citep[also see][]{Everalletal2019}. 
Such a large offset (the upper limit of estimates reported in the literature by
other authors) does indeed lead to tilt angles that are more
consistent with spherical alignment than obtained when using the
\citet{McMillan2018_simpleGaiaDR2distances_RVSsubset} distances.
Therefore it will be particularly important to pin-down, in future
\textit{Gaia} data releases, the amplitude of the parallax zero-point
as well as its local variations as these affect our ability to
constrain the mass distribution in our Galaxy.


\begin{acknowledgements}
The authors thank the anonymous referee whose insightful comments helped improving the quality of the manuscript. JH thanks Helmer Koppelman for helping in creating the dataset used in this work \citep{Koppelmanetal2018_Helmistreams}.
AH acknowledges financial support from a VICI grant from the Netherlands Organisation for Scientific Research, N.W.O. TdZ is grateful to the Kapteyn Astronomical Institute for the hospitality during his Blaauw Professorship.

This work has made use of data from the European Space Agency (ESA) mission {\it Gaia} (\url{https://www.cosmos.esa.int/gaia}), processed by the {\it Gaia} Data Processing and Analysis Consortium (DPAC, \url{https://www.cosmos.esa.int/web/gaia/dpac/consortium}). Funding for the DPAC has been provided by national institutions, in particular the institutions participating in the {\it Gaia} Multilateral Agreement.

We have also made use
of data from: (1) the APOGEE survey, which is part of Sloan Digital Sky Survey
IV. SDSS-IV is managed by the Astrophysical Research Consortium for the
Participating Institutions of the SDSS Collaboration (\url{http://www.sdss.org}).
(2) the RAVE survey (\url{http://www.rave-survey.org}), whose funding has
been provided by institutions of the RAVE participants and by their national
funding agencies. (3) the LAMOST survey (\url{www.lamost.org}), funded by the National
Development and Reform Commission. LAMOST is operated and managed by
the National Astronomical Observatories, Chinese Academy of Sciences. For
the analysis, the following software packages have been used: vaex \citep{Breddels&Veljanoski2018_vaex}, NumPy \citep{Oliphant2015_numpy}, matplotlib \citep{Hunter2007_matplotlib},
Jupyter Notebook \citep{Kluyveretal2016_jupyternotebook}, TOPCAT and STILTS \citep{Taylor2005_topcat,Taylor2006_stilts}.
\end{acknowledgements}



\bibliographystyle{aa}
\bibliography{bibliography}


\begin{appendix}

\section{Standard errors of sample (co)variances.} 
\label{App:Appendix_errorsofmoments}

To estimate the error on the inferred tilt angles from Method 1 of Sect. \ref{subsec:errordeconvolution} we propagate the errors of the relevant velocity moments from Eq. \ref{eq:tiltangle}. 

The error on a sample variance, $s^2$, can be estimated \citep[e.g.][]{Rao1973,Moodetal1974} by using
\begin{equation}
    \mathrm{var}(s^2) = \frac{1}{N} \left(  \mu_4 - \frac{N-3}{N-1}\mathrm{var}(v)^2 \right)
\end{equation}
for $N$ stars. Here, $\mu_4$ denotes the intrinsic $4^\mathrm{th}$ central moment and $s^2 = \frac{1}{N-1}\sum^N_{i=1} \left( v_i - \langle v \rangle \right)^2 $, for which $v_i$ is the relevant velocity component, either $v_R$ or $v_z$, of star $i$ and $\langle v \rangle$ its mean taken over all stars in the bin considered. The intrinsic velocity moments are estimated by their observed values, which is a good approximation given the relatively small errors in the data for the bins explored.

The error on a sample covariance $S_{xy}$ of $x$ and $y$ can be estimated (see \citet{Stuart&Ord1987} or \citet{Rose&Smith2002} for using mathStatica) by
\begin{equation}
    \mathrm{var}(S_{xy}) = \frac{1}{N} \left[ \mu_{22} - \frac{N-2}{N-1}\mathrm{cov}(x,y)^2 + \frac{1}{N-1}\mathrm{var}(x)\mathrm{var}(y)  \right] \, ,
\end{equation}
where $\mu_{22} = E[\{x-E(x)\}^2 \{y-E(y)\}^2]$ for $E$ denoting the expectation value. We have defined $S_{xy} = \frac{1}{N-1}\sum^N_{i=1} ( x_i - \langle x \rangle ) \, ( y_i - \langle y \rangle ) $. In our application $x$ is replaced for $v_R$ and $y$ for $v_z$. The intrinsic moments are again estimated by taking the equivalent moments directly from the observed velocity distribution.


\begin{figure*}[t]
  \centering
  \includegraphics[width=1.0\textwidth]{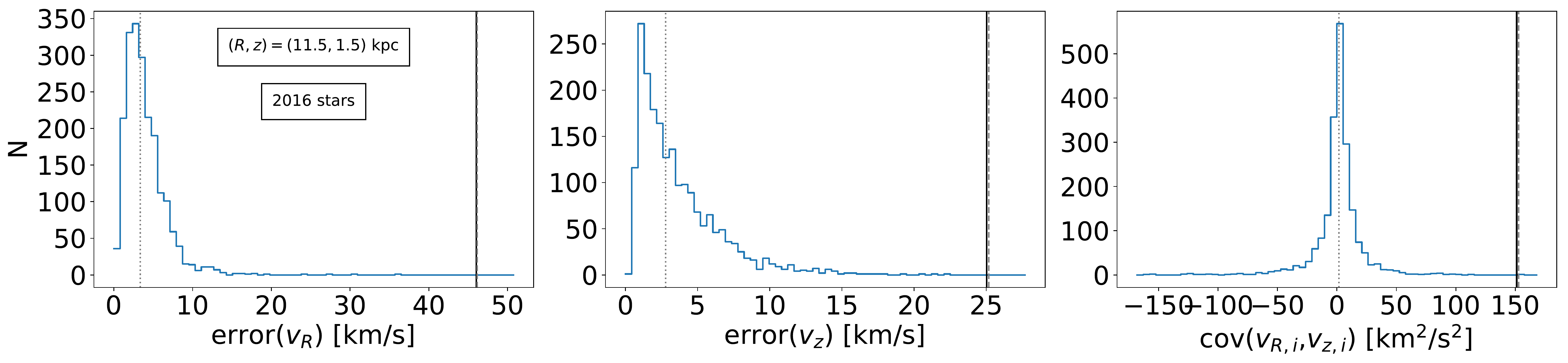}
  \caption{Error distributions for the bin at $R=11.5$~kpc and $z=1.5$~kpc for the different velocity components: $v_R$ (left), $v_z$ (middle), and its covariance (right). The corresponding medians of the error distributions are shown by the vertical grey dotted lines. The vertical grey dashed lines indicate the values of the velocity moments taken from the data directly (i.e. not accounting for the errors). 
  The black vertical solid lines lines show the recovered intrinsic velocity moment from Method 1 (see Sect. \ref{sec:methods}).
  Even at this bin, which still contains $2,016$ stars, the impact of the measurement errors on the recovered velocity moments is relatively small.} 
  \label{fig:exampleerror} 
\end{figure*}%

As an example for Sect. \ref{subsec:errordeconvolution} we show in Fig. \ref{fig:exampleerror} the error distributions for the bin at $R=11.5$~kpc and $z=1.5$~kpc. This bin is near the edge of the volume investigated, but still contains $2,016$ stars. The vertical grey dashed lines indicate the values of the velocity moments that would be derived by using the data directly (i.e. not accounting for the errors). The medians of the error distributions are indicated by the vertical grey dotted lines. The recovered intrinsic velocity moments from Method 1 are visualised by the vertical black solid lines (as here, these usually coincide with the vertical grey dashed lines).  
Thus, even for this outer bin, the effects of measurement errors are relatively small.


\section{The impact of a systematic parallax offset on the recovered tilt angles.}
\label{App:Appendix_parallaxoffset}
Here we explain how a systematic parallax offset can affect the inferred tilt angles. For this purpose, we now only consider the $(x,z)$-plane and we assume that all parallaxes are shifted by the same offset $\Delta \varpi = -0.029$~mas. 

For Galactic longitude $l$ and latitude $b$ the $(U,V,W)$-velocities in km/s can be computed the usual way \citep{Johnson&Soderblom1987, Bovy2011}:
\begin{equation}
\label{eq:UVW}
\begin{pmatrix} 
U \\ 
V \\ 
W  
\end{pmatrix}
=
\begin{pmatrix} 
\cos(l) \cos(b) & -\sin(l) & -\cos(l) \sin(b) \\ 
\sin(l) \cos(b) & \cos(l) & -\sin(l) \sin(b) \\ 
\sin(b) & 0 & \cos(b)  
\end{pmatrix}
\begin{pmatrix} 
v_\mathrm{los} \\ 
\frac{k}{\varpi} \mu_{l^\star} \\ 
\frac{k}{\varpi} \mu_{b}  
\end{pmatrix}
\, .
\end{equation}
Here, $\mu_{l^\star}=\mu_l \cos(b)$ and $\mu_b$ denote the proper motions in mas/yr in the direction of $l$ and $b$, respectively, $\varpi$ is the parallax in mas, and $k=4.74047 \frac{\textrm{km/s}}{\textrm{kpc mas/yr}}$ (assuming a Julian year). 

When only considering an error in the parallaxes the `observed' velocities are affected as:
\begin{equation}
\label{eq:UVW0toUVW1}
\begin{pmatrix} 
U \\ 
V \\ 
W  
\end{pmatrix}_\mathrm{1}
=
\begin{pmatrix} 
U \\ 
V \\ 
W  
\end{pmatrix}_\mathrm{0}
+
\left.
\frac{\partial}{\partial \varpi} \begin{pmatrix} 
U \\ 
V \\ 
W  
\end{pmatrix}
\right \vert_\mathrm{0}
\Delta \varpi
+
O(\Delta \varpi^2) 
\, .
\end{equation}
Subscript 0 denotes the true position and velocities, subscript 1 the `observed' quantities. Furthermore:
\begin{equation}
\label{eq:derivative}
\frac{\partial}{\partial \varpi} \begin{pmatrix} 
U \\ 
V \\ 
W  
\end{pmatrix}
=
- \frac{1}{\varpi}
\begin{pmatrix} 
U - \cos(l) \cos(b) \, v_\mathrm{los} \\ 
V - \sin(l) \cos(b) \, v_\mathrm{los}\\ 
W - \sin(b) \, v_\mathrm{los} \qquad \; 
\end{pmatrix} \, ,
\end{equation} 
and:
\begin{equation}
    v_\mathrm{los} = \cos(b) \cos(l) \, U + \cos(b) \sin(l) \, V + \sin(b) \, W \, .
\end{equation}

Let us now define the tilt angle $\delta$ as:
\begin{equation}
\label{eq:tiltangle_UW}
    \tan(2 \delta) = \frac{2 \mathrm{cov}(U, W)}{\mathrm{var}(U) - \mathrm{var}(W)} \, .
\end{equation}
In a steady state axisymmetric system $\langle v_R \rangle = \langle v_z \rangle =0$.
Therefore, at the $(x,z)$-plane $\langle U \rangle = \langle W \rangle = 0$, and thus $\textrm{var}(U) = \langle U^2 \rangle $, $\textrm{var}(W) = \langle W^2 \rangle $, and $\textrm{cov}(U, W) = \langle UW \rangle$.
For $l=0^{\circ}$ and $l=180^{\circ}$ we also notice that $U=-v_R$ and $W=v_z$ , and therefore that $\delta = - \gamma$. In the remainder of this Appendix we refer to $\delta$ when we use `tilt angle' (unless stated otherwise).

Plugging Eq. \ref{eq:UVW0toUVW1} up to first order in $\frac{\Delta \varpi}{\varpi_0}$ into Eq. \ref{eq:tiltangle_UW} we get:
\begin{equation}
\label{eq:tiltangle0_to_tiltangle1}
    \tan(2 \delta_1) \simeq \frac{2\langle U_0 W_0 \rangle + \epsilon_\mathrm{A}}{\langle U^2_0 \rangle - \langle W^2_0 \rangle + \epsilon_\mathrm{B}} \, ,
\end{equation}
in which:
\begin{equation}
\label{eq:epsilonAB}
\begin{aligned}
  \epsilon_A &= \left[ \pm \left(\langle U^2_0 \rangle + \langle W^2_0 \rangle \right) \sin(2b) - 2\langle U_0 W_0 \rangle \right] \left(\frac{\Delta \varpi}{\varpi_0}\right) \\
  \epsilon_B &= 2 \left[\langle W_0^2 \rangle \cos^2(b) - \langle U_0^2 \rangle \sin^2(b)  \right] \left(\frac{\Delta \varpi}{\varpi_0}\right)\, ,
\end{aligned} 
\end{equation} 
where $\pm$ holds for $l \in \begin{Bmatrix} 0^\circ \\ 180^\circ \end{Bmatrix}$.

To further explore the effect of a shift in the parallaxes we now
investigate what would happen to the tilt angles in two different
cases of alignment: spherical alignment and cylindrical alignment.

We start by rewriting Eq. \ref{eq:tiltangle0_to_tiltangle1} into the form of
\begin{equation}
\label{eq:tiltangle0_to_tiltangle1_goal}
\begin{aligned}
    \delta_1 &= \frac{1}{2} \arctan \left[ \left( 1+x \right) \tan(2\delta_0) \right] \\
    \delta_1 &= \delta_0 + \frac{1}{4} \sin(4 \delta_0) \, x + O(x^2) \\
    \Delta \delta &\simeq \frac{1}{4} \sin(4 \delta_0) \, x \, .
\end{aligned} 
\end{equation}
We then get:
\begin{equation}
\label{eq:tiltangle0_to_tiltangle1_rewrited}
\begin{aligned}
    \tan(2 \delta_1) &\simeq \frac{2\langle U_0 W_0 \rangle}{\langle U^2_0 \rangle - \langle W^2_0 \rangle} \left[ \frac{1-\epsilon_\mathrm{C}}{1 - \epsilon_\mathrm{D}} \right] \\
    \tan(2 \delta_1) &\simeq \tan(2 \delta_0) \left[ \frac{1-\epsilon_\mathrm{C}}{1 - \epsilon_\mathrm{D}} \right] \, ,
\end{aligned}
\end{equation}
in which:
\begin{equation}
\label{eq:epsilonCD}
\begin{aligned}
  \epsilon_\mathrm{C} &= \left[1 \mp \left( \frac{\langle U^2_0 \rangle + \langle W^2_0 \rangle}{2\langle U_0 W_0 \rangle} \right) \sin(2b)\right] \left(\frac{\Delta \varpi}{\varpi_0}\right) \\
  \epsilon_\mathrm{D} &= 2 \left[ \frac{\langle U_0^2 \rangle \sin^2(b) - \langle W_0^2 \rangle \cos^2(b)}{\langle U^2_0 \rangle - \langle W^2_0 \rangle} \right] \left(\frac{\Delta \varpi}{\varpi_0}\right) \, .
\end{aligned} 
\end{equation} 
Then, under the assumptions that $\left \vert \epsilon_C \right \vert\ll1$ and $\left \vert \epsilon_D \right \vert\ll1$, we get:
\begin{equation}
\label{eq:tiltangle0_to_tiltangle1_x}
    x \simeq \epsilon_\mathrm{D} - \epsilon_\mathrm{C} \, .
\end{equation}
We highlight the effects for four different latitudes:
\begin{equation}
\label{eq:tiltangle0_to_tiltangle1_difb}
\begin{aligned}
  &b=0^\circ: \qquad       &&x= - \left(\frac{\Delta \varpi}{\varpi_0}\right) \left[ \frac{\langle U^2_0 \rangle + \langle W^2_0 \rangle}{\langle U^2_0 \rangle - \langle W^2_0 \rangle} \right] \\
  &\vert b \vert =90^\circ: \qquad     &&x= + \left(\frac{\Delta \varpi}{\varpi_0}\right) \left[ \frac{\langle U^2_0 \rangle + \langle W^2_0 \rangle}{\langle U^2_0 \rangle - \langle W^2_0 \rangle} \right] \\
  &b=+45^\circ: \qquad     &&x= \pm \left(\frac{\Delta \varpi}{\varpi_0}\right) \left[ \frac{\langle U^2_0 \rangle + \langle W^2_0 \rangle}{2\langle U_0 W_0 \rangle} \right] \\
  &b=-45^\circ: \qquad     &&x= \mp \left(\frac{\Delta \varpi}{\varpi_0}\right) \left[ \frac{\langle U^2_0 \rangle + \langle W^2_0 \rangle}{2\langle U_0 W_0 \rangle} \right] \, .
\end{aligned} 
\end{equation} 
Since the velocity ellipse is mostly non-tilted ($\delta_0 = 0^\circ$) at the Galactic midplane the inferred tilt angles at $b=0^\circ$ are not affected by an error in the parallax. Geometrically this is not surprising since, at $b=0^\circ$, the $U$-component of the velocities are not affected. The $W$-velocities are only inflated and do not change the tilt angle. However, if $\delta_0 \neq 0^\circ$, then the term between the square brackets becomes larger than one, since for typical values of the velocity moments at the midplane $\sigma(v_R) > \sigma(v_z)$ \citep[see e.g.][]{GaiaDR2_disckinematis_Katzetal2018}. The inferred tilt angle is therefore steeper (more positive if $\delta_0>0^\circ$ and more negative if $\delta_0<0^\circ$). At $\vert b \vert =90^\circ$, the effect is reversed and the tilt angle becomes shallower (less positive if $\delta_0>0^\circ$ and less negative if $\delta_0<0^\circ$) due to the parallax offset. For the case of spherical alignment the relation $\tan(2\delta_0) = \tan(2\theta)$ can be applied. 

The approximations used so far fail for $\langle U_0 W_0 \rangle \simeq 0$, since then $|\epsilon_\mathrm{C}| \not \ll 1 $, and for $\langle U^2_0 \rangle \simeq \langle W^2_0 \rangle$, since then $|\epsilon_\mathrm{D}| \not \ll 1 $, and thus $|x| \not \ll 1$. In the case of cylindrical alignment ($\delta_0 = \langle U_0 W_0 \rangle=0$) and for
$\left \vert \epsilon_\mathrm{D} \right \vert \ll 1$ 
we get\footnote{If, hypothetically, both $\langle U^2_0 \rangle=\langle W^2_0 \rangle$ and $\langle U_0 W_0 \rangle=0$, then $\tan(2 \delta_1) = \pm \tan(2 b)$.}:
\begin{equation}
\label{eq:tiltangle0_to_tiltangle1_cylindrical}
    \delta_1 \simeq \pm \frac{1}{2} \sin(2b) \left( \frac{\langle U^2_0 \rangle + \langle W^2_0 \rangle}{\langle U^2_0 \rangle - \langle W^2_0 \rangle} \right) \left(\frac{\Delta \varpi}{\varpi_0} \right)  \, ,
\end{equation}
where we used that $\tan(2 \delta_1) \simeq 2\delta_1$ for small deviations around $\delta_1=0^\circ$. This means that at $l=0$ ($l=180^\circ$) and for $\sigma(v_R) > \sigma(v_z)$ the tilt angles appear to be negative (positive) for $b>0^\circ$, and positive (negative) for $b<0^\circ$.

\begin{figure*}[t]
  \centering
  \includegraphics[width=1.0\textwidth]{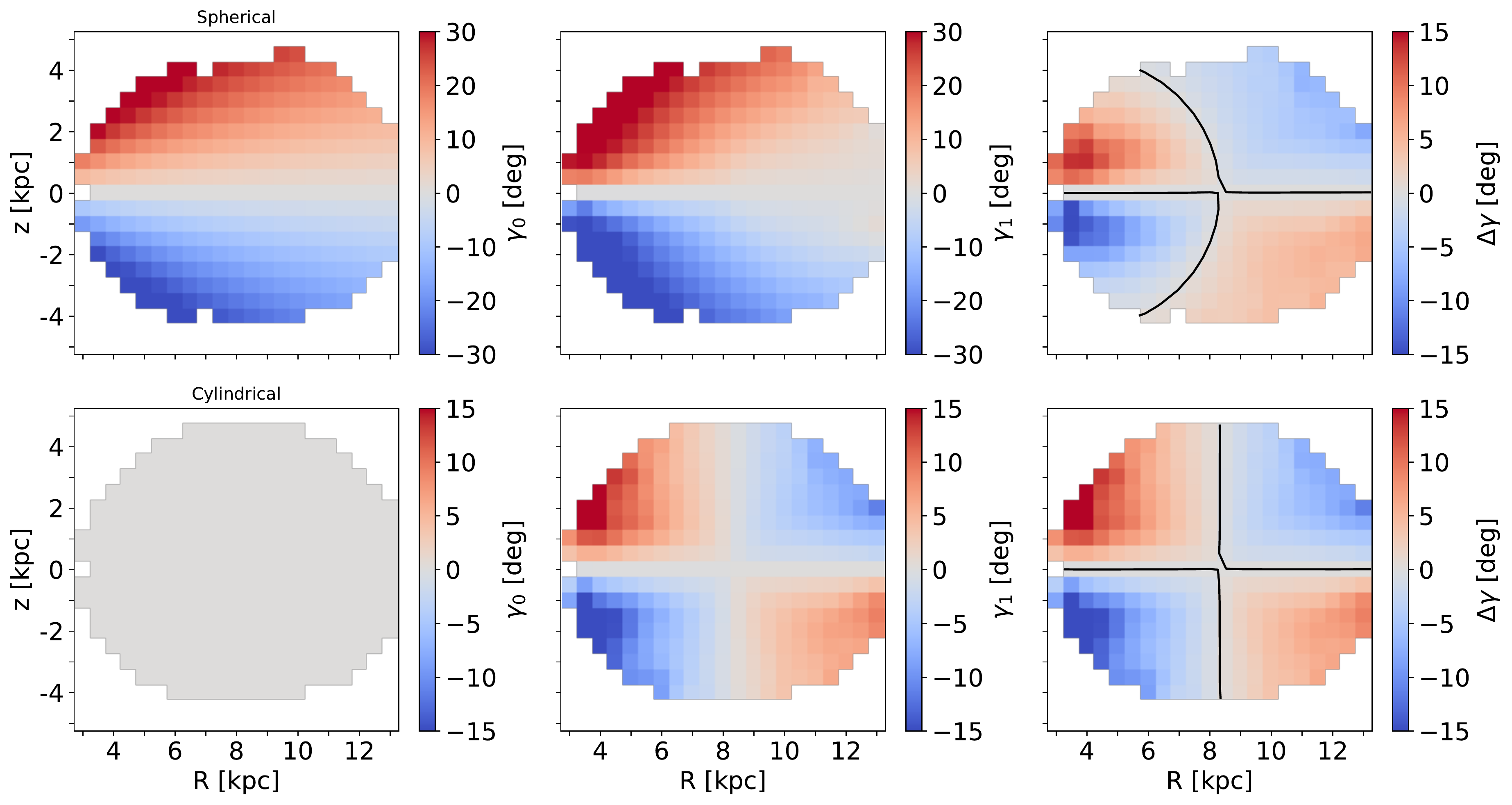}
  \caption{The effect of a constant shift in the parallaxes of the
    stars ($\Delta \varpi = -0.029$~mas) on the tilt angle $\gamma$, as
    measured in Galactocentric cylindrical coordinates for different
    types of intrinsic alignment. The left columns show intrinsic tilt
    angles $\gamma_0$ as a function of $R$ and $z$. The middle columns
    show the tilt angles $\gamma_1$ computed from the `observed'
    velocity moments. The right
    column shows $\Delta \gamma = \gamma_1 - \gamma_0$. Be aware of the different colourbar ranges.
    In the top panels we set the velocity covariances such that the input alignment is spherical.
    In the bottom panels the input alignment is cylindrical.
    Black contours denote regions where the tilt angle is not affected, i.e. $\Delta \gamma =0^\circ$. For spherical
    alignment this is expected to be the case on the line passing through the Galactic centre and the position of the Sun, thus along $z\approx 0$~kpc, and on the circle that goes
    through the Galactic centre and the position of the Sun. For cylindrical
    alignment this is expected to occur at both $z=z_{\odot} \approx 0$~kpc and
    $R=R_\odot$.}
  \label{fig:analytic_parallax_effect} 
\end{figure*}

We have inserted the relevant Galactic velocity dispersions as a function of $R$ and $z$ and set the covariance term such that there is either spherical or cylindrical alignment throughout the extent of the dataset. We find that the tilt angles are affected very similarly. This is visualised in Fig. \ref{fig:analytic_parallax_effect} (recall that $\gamma = - \delta$ since we here consider $l=0^{\circ}$ and $l=180^{\circ}$ only). We therefore think that our test performed in Sect. \ref{subsec:parallaxoffset} is realistic, even though the intrinsic tilt angles of the GUMS catalogue are more or less cylindrically aligned.

Besides the fact that the orientation of the velocity ellipse changes due to the parallax offset, obviously the stars under consideration also move in position. Thus, in fact a sample of stars with tilt angle $\delta_0$ at parallax $\varpi_0$ gets `observed' at $\varpi_1$ with tilt angle $\delta_1$. We have not taken this effect into account in the analytic description from this Appendix.

\end{appendix}

\end{document}